\documentclass[11pt]{article}
\usepackage{setspace}
\usepackage{cite}
\usepackage{amsmath,amssymb,amsfonts}
\usepackage{amsmath}
\usepackage{graphicx}
\usepackage{textcomp}
\usepackage{xcolor}
\usepackage{balance} 
\usepackage{array}
\usepackage{float} 
\usepackage{scalerel}
\usepackage{multicol}
\usepackage{esdiff}
\usepackage[hidelinks]{hyperref}
\usepackage[capitalise]{cleveref}
\usepackage{cases}
\usepackage{lipsum}
\usepackage{balance}
\usepackage{mathtools}
\usepackage{comment}
\usepackage{algorithm}
\usepackage{algpseudocode}
\usepackage{subcaption}
\usepackage{caption}
\usepackage{graphicx}
\usepackage{placeins}
\usepackage{textcase}
\usepackage{tikz}
\usetikzlibrary{arrows.meta,positioning}
\usepackage[a4paper,margin=1in]{geometry}
\usepackage{microtype}
\usepackage{adjustbox}
\usepackage{booktabs}
\usepackage{array}
\DeclareCaptionFont{ninept}{\fontsize{9pt}{10.8pt}\selectfont}
\usepackage[hidelinks]{hyperref}

\usepackage[normalem]{ulem}

\title{A Distributed Quantum Approximate Optimization Algorithm Simulator for Engineering Design Optimization}
\author{Ali Rajabi, Milad Hasanzadeh, Amin Kargarian}

\begin{document}

\maketitle

\begin{abstract}
\noindent This paper presents a Qiskit-compatible distributed quantum approximate optimization algorithm (DQAOA) simulator for the general class of quadratic unconstrained binary optimization (QUBO) problems that arise in a wide range of engineering design applications. \uline{The simulator is open source and publicly available on the \href{https://sites.google.com/site/aminkargarian/home}{RAISE LAB website} and \href{https://github.com/LSU-RAISE-LAB/}{RAISE LAB GitHub repository}}, where users can access it and read the accompanying README documentation for installation, input formatting, configurable parameters, and example workflows. The package is motivated by the lack of a general, reusable simulator that can consistently solve QUBO instances across different QAOA execution modes. The developed simulator package supports monolithic QAOA on a single quantum processing unit (QPU) and distributed QAOA across a user-specified number of QPUs with configurable capacities. The number of logical qubits is determined by the input QUBO size, allowing users to study different problem sizes while adjusting the multi-QPU setting. The framework canonicalizes the QUBO model, maps it to a cost Hamiltonian, allocates variables across QPUs, identifies local and cross-QPU couplings, and constructs the corresponding circuits within one workflow. Runtime optimization strategies, including parameterized circuit reuse, objective reuse at fixed depth, batched evaluations, and parallel multi-start execution, are applied to reduce repeated software overhead and facilitate QUBO solution and comparison. A graphical user interface developed by Streamlit is also provided to make the simulator easier to use, allowing users to enter or upload QUBO instances, configure solver settings, run selected modes, and visualize solution-quality metrics without directly editing Python scripts. The package is also demonstrated in a power generation unit commitment application. In this application, brute force, monolithic QAOA, and distributed QAOA modes recover the same generation unit commitment bitstring and operating cost. Results across multiple case studies are consistent with classical monolithic QAOA references regarding optimal bitstrings and costs. The staged runtime analysis shows substantial runtime reduction across implementation stages, while DQAOA remains more demanding because cross-QPU couplings require explicit remote operations.
\end{abstract}


\section{Introduction}

Binary optimization problems arise in many engineering applications where decisions must be made under practical operating, planning, and design constraints \cite{glover2018tutorial}. Examples include air traffic optimization, gas network expansion planning, water distribution network management, power grid operation and planning, resource allocation in food systems, and finance optimization problems \cite{moncayo2025quantum, bertsimas1998air, borraz2016convex, savic1997genetic, li2021optimising, markowitz2008portfolio, manousakis2012taxonomy}. In power systems, binary and mixed-integer decision structures appear in applications such as phasor measurement unit placement, distribution-system reconfiguration, grid operation, and planning under practical network and device constraints \cite{fathizadan2022holistic,torkaman2021multi,monteiro2020electric,liao2021review,amani2026learning, amani2026event}. These problems have traditionally been addressed using classical optimization methods. Classical approaches for binary optimization commonly rely on exact and heuristic solution methods, including branch-and-bound, branch-and-cut, decomposition-based algorithms, and metaheuristic search methods \cite{wolsey2020integer,glover2003handbook}. To improve scalability, researchers have also studied parallel implementations of classical optimization algorithms and machine-learning-assisted approaches for combinatorial optimization \cite{eckstein1994parallel,bengio2021machine}. In these approaches, the solver must account for the fact that each decision variable can take only two possible values, 0 or 1, which creates a combinatorial search space. Although these methods are widely used in engineering optimization and can provide reliable solutions for many practical instances, their computational effort can increase rapidly as the number of binary variables and coupling constraints grows. As a result, classical solvers may face significant challenges for large-scale or strongly coupled binary optimization problems.

More recently, quantum computing has been explored as an alternative computational direction for optimization problems \cite{hasanzadeh2026survey}. However, an optimization problem cannot be used directly by a quantum algorithm in its original engineering formulation. It must first be expressed in a mathematical form that can be encoded into a quantum system. For binary optimization problems, this is commonly done by rewriting the objective in a form that can be mapped to an Ising Hamiltonian, whose energy levels correspond to the objective values of the binary solutions \cite{lucas2014ising}. Quadratic Unconstrained Binary Optimization (QUBO) provides a convenient algebraic representation for this purpose because it describes the objective using binary variables and pairwise quadratic interactions. Therefore, QUBO serves as an important bridge between classical binary optimization models and quantum optimization frameworks \cite{glover2019quantum}.

In a QUBO model, the objective is written in terms of binary variables via linear and quadratic terms that explicitly represent the coupling among decision variables \cite{glover2018tutorial}. This representation is especially important because many combinatorial optimization problems can be expressed in equivalent Ising or QUBO form \cite{lucas2014ising}. As a result, QUBO has become a standard modeling framework for connecting classical binary optimization problems to quantum optimization algorithms \cite{glover2018tutorial}.

The connection between QUBO and the Ising Hamiltonian has made quantum optimization methods increasingly attractive for binary decision problems. The motivation is not that quantum methods have already replaced classical optimization in practice, but that they provide a different computational framework for representing and exploring large binary search spaces. In particular, quantum circuits can prepare superposition states that encode many candidate solutions at once, while entanglement can capture correlations among variables during the evolution of the quantum state. For this reason, quantum algorithms have attracted growing interest for combinatorial optimization, especially in the near-term setting \cite{abbas2024challenges}.

Several quantum and hybrid quantum-classical approaches have been studied for optimization problems, including quantum annealing, adiabatic quantum computing, the Variational Quantum Eigensolver (VQE), and the Quantum Approximate Optimization Algorithm (QAOA). Quantum annealing uses quantum fluctuations to search for low-energy configurations of an optimization Hamiltonian \cite{kadowaki1998quantum}. Adiabatic quantum computing solves problems by slowly evolving a quantum system toward a final Hamiltonian whose ground state encodes the solution \cite{farhi2000quantum}. VQE uses a parameterized quantum circuit together with a classical optimizer to minimize the expectation value of a Hamiltonian \cite{peruzzo2014variational}. Among these approaches, QAOA is one of the most widely studied methods for combinatorial optimization because it is designed specifically for discrete optimization problems and can be implemented as a gate-based variational algorithm \cite{farhi2014quantum}. In QAOA, the optimization objective is encoded into a quantum cost Hamiltonian, and a parameterized quantum circuit is constructed by alternating cost and mixer operations. A classical optimizer then updates the circuit parameters to guide the measured quantum state toward low-cost solutions \cite{zhou2020quantum}. Since its introduction, QAOA has been studied from several directions, including extensions of the original framework, performance analysis, and implementation on near-term devices. One important extension is the Quantum Alternating Operator Ansatz, which generalizes the original QAOA construction to a broader class of optimization problems \cite{hadfield2019quantum}. Broader reviews of QAOA, its variants, and its implementation challenges are presented in \cite{blekos2024review}.

Although QAOA has emerged as a promising approach for combinatorial optimization, its practical performance varies across problem instances. Empirical studies have shown that the achieved performance can vary substantially with both the problem structure and the chosen circuit depth \cite{lotshaw2021empirical}. This has motivated several efforts to improve the variational search process. For example, Conditional Value at Risk (CVaR) was proposed to place greater emphasis on the lowest-cost sampled solutions during training \cite{barkoutsos2020improving}. Warm-start methods were also introduced to incorporate classical information into the initialization of the variational search \cite{egger2021warm}. In addition, the reuse and transfer of QAOA parameters across related instances have been studied as a way to improve practical optimization behavior \cite{galda2021transferability,shaydulin2023parameter}. Parameter-setting strategies for weighted QAOA problems were investigated in \cite{sureshbabu2024parameter}. Even with these improvements, low-depth QAOA can still be difficult to train effectively \cite{rajakumar2024trainability}. These observations are consistent with broader challenges in near-term quantum optimization \cite{abbas2024challenges}.

Another practical limitation of monolithic QAOA is that it assumes all qubits and all required interactions can be accommodated within a single quantum processing unit (QPU). As the size of the binary optimization problem grows, this single-QPU assumption can become difficult to maintain due to hardware limits on the number of qubits and operations supported by a single QPU. In this context, decomposition ideas for QUBO already indicate the value of partitioning a large problem when direct solution becomes difficult \cite{guerreschi2021solving}. From the quantum-computing side, distributed quantum computing has therefore attracted attention as a way to execute larger circuits across multiple QPUs \cite{diadamo2021distributed}. The transition from single-QPU systems to distributed quantum architectures has also been discussed in recent review work \cite{barral2025review}. At the circuit level, this requires deciding how to carry out operations involving qubits on different QPUs \cite{sundaram2022distribution}. It also requires deciding how qubits should be assigned to the available QPUs, since this assignment directly affects how many interactions must cross QPU boundaries \cite{mao2023qubit}.

These developments have motivated distributed versions of QAOA. A local-to-global distributed QAOA framework for pseudo-Boolean optimization was introduced in \cite{yue2023local}. Distributed QAOA on a quantum-centric supercomputing architecture was studied in \cite{kim2024distributed}. Noise-aware distributed QAOA on near-term quantum hardware was reported in \cite{chen2024noise}. Related distributed variational quantum methods have also been explored in power-system applications, including a distributed Variational Quantum Eigensolver framework for unit commitment \cite{hasanzadeh2026distributed}. Together, these studies show that distributed QAOA is a promising direction to extend quantum optimization beyond the single-QPU setting. At the same time, they also show that practical distributed execution depends not only on the QAOA formulation itself, but also on variable allocation, cross-QPU interaction implementation, and the software overhead introduced by repeated circuit execution.

Despite these advances, the existing literature is still largely focused on specific problem settings, hardware assumptions, or execution models. As a result, there remains a need for a general simulator and software framework that can accept QUBO problems directly and solve them in both monolithic and distributed QAOA settings within a single consistent workflow. Such a tool is important because it allows users to study the same QUBO instance under different execution modes, compare the solution quality and runtime behavior, and evaluate the effect of QPU partitioning without rebuilding a separate workflow for each case.

Motivated by this gap, this paper develops an open-source, Qiskit-compatible distributed quantum approximate optimization algorithm (DQAOA) simulator for QUBO problems that arise in a wide range of engineering design applications. The simulator is implemented using Qiskit and supports multiple solver modes, including classical reference modes, monolithic QAOA on a single QPU, and distributed QAOA across multiple QPUs. In the monolithic QAOA mode, all variables and interactions are handled by a single QPU. In the distributed QAOA mode, the user can specify an arbitrary number of QPUs and their capacities, and the package assigns the QUBO variables across the available QPUs accordingly. The number of logical qubits is determined by the number of binary variables in the user-provided QUBO instance, so users can study problems of different sizes by changing the QUBO input. This makes the package suitable for analyzing how QUBO size, QPU capacity, variable allocation, and cross-QPU interactions affect QAOA-based optimization.

The simulator is evaluated on multiple QUBO benchmark cases with different problem sizes, QPU counts, QPU capacity assignments, variable-allocation strategies, training-shot settings, optimizer-iteration limits, and final sampling shot settings. These tests allow the framework to be examined under a wide range of solver configurations rather than only under a single fixed setup. In addition to the QUBO benchmarks, the package is also tested on a unit commitment (UC) application in power systems. In that case, the DQAOA package is used as the QUBO solver for the binary commitment block inside a three-block ADMM framework. This application demonstrates how the developed simulator can be embedded in a larger engineering optimization workflow rather than being limited to isolated QUBO examples. In addition to the Python callable interface, the package includes a Streamlit-based graphical user interface that allows users to configure QUBO inputs, QAOA settings, QPU capacities, solver modes, and output visualizations through an interactive web-based dashboard.

The proposed framework begins with a canonical QUBO representation, maps it to the corresponding cost Hamiltonian, allocates variables across QPUs, identifies local and cross-QPU couplings, and constructs monolithic and distributed quantum circuits within one consistent workflow. The paper also studies runtime optimization strategies that reduce repeated circuit construction and transpilation overhead, reduce optimization overhead, and improve execution efficiency during repeated shot-based evaluations. These strategies help facilitate the practical solution and comparison of QUBO problems using the proposed framework. \uline{The simulator is open source and publicly available on the \href{https://sites.google.com/site/aminkargarian/home}{RAISE LAB website} and \href{https://github.com/LSU-RAISE-LAB/}{RAISE LAB GitHub repository}}~\cite{RAISELABWebsite,RAISELABGitHub}, where users can access it and read the accompanying README documentation for installation, input formatting, configurable parameters, and example workflows.
The numerical results show that the monolithic and distributed modes remain consistent with classical reference solutions in terms of the optimal bitstring and optimal cost across the tested QUBO cases. The runtime comparison clarifies the practical effect of the implementation refinements introduced in the software, while the UC application illustrates the use of the package as a QUBO solver module inside a larger engineering optimization workflow.

\section{Preliminaries}

This section presents the background needed to develop DQAOA and establish the problem formulation used throughout the paper. We describe the QUBO representation and its mapping to the corresponding quantum cost Hamiltonian. We review the standard QAOA algorithm as the monolithic baseline. We explain the main concepts of distributed quantum execution, which provide the basis for the multi-QPU framework developed in Section \ref{DQAOAFramework}.

\subsection{QUBO Formulation}

QUBO provides a compact representation for a broad class of binary optimization problems and is widely used in quantum optimization because it can be mapped to Ising-type objective functions \cite{glover2018tutorial,lucas2014ising}. In its standard form, a QUBO problem seeks a binary vector $\mathbf z\in\{0,1\}^n$, where $n$ denotes the number of binary decision variables, that minimizes a quadratic objective function:

\begin{equation}
\min_{\mathbf z\in\{0,1\}^n} \; f(\mathbf z)=\mathbf z^\top Q \mathbf z + c.
\label{eq:qubo_standard}
\end{equation}
$Q\in\mathbb R^{n\times n}$ is a real-valued coefficient matrix and $c\in\mathbb R$ is a constant offset. Because the decision variables are binary, the objective can be written equivalently as a constant term plus linear and pairwise quadratic contributions. This modeling flexibility makes QUBO a convenient interface between classical combinatorial optimization and quantum optimization methods.

We specify each problem instance by a dense quadratic matrix $H\in\mathbb R^{n\times n}$, a linear coefficient vector $\mathbf f\in\mathbb R^n$, and a constant term $c_0\in\mathbb R$. Using the binary decision vector $\mathbf z\in\{0,1\}^n$, the objective is written as
\begin{equation}
F(\mathbf z)=c_0+\mathbf f^\top \mathbf z+\mathbf z^\top H \mathbf z.
\label{eq:qubo_input_form}
\end{equation}
To obtain the canonical representation used throughout the paper, the quadratic term is expanded into diagonal and off-diagonal parts:
\begin{equation}
\mathbf z^\top H \mathbf z
=
\sum_{i=1}^n H_{ii} z_i^2
+
\sum_{i=1}^n \sum_{j=i+1}^n (H_{ij}+H_{ji}) z_i z_j.
\label{eq:qubo_expand_H}
\end{equation}
Since $z_i^2=z_i$ for binary variables, the diagonal terms can be absorbed into the linear part of the objective. The canonical coefficients are defined as
\begin{equation}
\ell_i=f_i+H_{ii}, \qquad i=1,\dots,n,
\label{eq:canonical_linear}
\end{equation}
and
\begin{equation}
b_{ij}=H_{ij}+H_{ji}, \qquad 1\le i<j\le n.
\label{eq:canonical_quadratic}
\end{equation}
The resulting canonical QUBO form is therefore
\begin{equation}
F(\mathbf z)
=
c_0
+
\sum_{i=1}^n \ell_i z_i
+
\sum_{i=1}^n \sum_{j=i+1}^n b_{ij} z_i z_j.
\label{eq:qubo_canonical_work}
\end{equation}

This representation is adopted for three reasons. First, it removes redundancy by storing each pairwise interaction only once through the restriction $i<j$. Second, it separates single-variable contributions from pairwise couplings, which simplifies both objective evaluation and the later construction of the QAOA cost Hamiltonian. Third, because the pairwise terms are written explicitly, the interaction structure of the problem becomes easier to analyze in the distributed setting considered in this work. 

\subsection{Mapping QUBO to Quantum Cost Hamiltonian}

To solve QUBO \eqref{eq:qubo_canonical_work} using QAOA, the objective function must be encoded into a quantum cost Hamiltonian whose computational basis eigenvalues match the objective values of the bitstrings. Using the standard binary-to-qubit mapping,
\begin{equation}
\hat z_i=\frac{I-Z_i}{2},
\label{eq:binary_to_qubit}
\end{equation}
where $I$ is the identity operator and $Z_i$ is the Pauli-$Z$ operator acting on qubit $i$, the canonical QUBO can be mapped to the cost Hamiltonian as
\begin{equation}
H_C
=
c_0 I
+
\sum_{i=1}^n \ell_i \frac{I-Z_i}{2}
+
\sum_{i=1}^n \sum_{j=i+1}^n b_{ij}\frac{(I-Z_i)(I-Z_j)}{4}.
\label{eq:cost_hamiltonian_raw}
\end{equation}
This Hamiltonian is diagonal in the computational basis, and each basis state $|\mathbf z\rangle$ is an eigenstate of $H_C$ with eigenvalue equal to $F(\mathbf z)$. Therefore, minimizing QUBO \eqref{eq:qubo_canonical_work} is equivalent to identifying the ground state of the corresponding cost Hamiltonian \eqref{eq:cost_hamiltonian_raw}.

By expanding the products in quadratic terms, the Hamiltonian can be written in the standard Pauli form
\begin{equation}
H_C=\alpha I+\sum_{i=1}^n h_i Z_i+\sum_{i=1}^n\sum_{j=i+1}^n J_{ij} Z_i Z_j,
\label{eq:cost_hamiltonian_pauli}
\end{equation}
where $\alpha$ is a constant offset, $h_i$ are effective single-qubit coefficients, and $J_{ij}$ are pairwise $ZZ$ couplings determined by QUBO coefficients. The identity term contributes only a global phase to the cost unitary and does not affect the optimizer, whereas the $Z_i$ and $Z_iZ_j$ terms define the nontrivial problem-dependent phase evolution.

This Pauli representation directly defines the QAOA cost unitary, or phase-separation operator, as
\begin{equation}
U_C(\gamma)=e^{-i\gamma H_C},
\label{eq:qaoa_cost_unitary}
\end{equation}
where $\gamma$ is a variational parameter. Since all terms in $H_C$ are diagonal and mutually commuting, the cost unitary can be implemented as a sequence of commuting single-qubit and two-qubit phase operations. In particular, linear coefficients contribute single-qubit $Z$-phase rotations, while quadratic coefficients contribute pairwise $ZZ$-type interactions.

This mapping is fundamental to the DQAOA framework presented in Section \ref{DQAOAFramework}. Once a QUBO instance is converted into its corresponding cost Hamiltonian, the set of pairwise couplings determines the interaction structure of the problem. In the monolithic case, all such interactions are executed within a single QPU. In the distributed case, however, the same $Z_iZ_j$ terms must be examined relative to the variable-to-QPU assignment to determine whether they correspond to local or cross-QPU couplings. Thus, the Hamiltonian representation serves as the bridge between the QUBO formulation and the distributed QAOA circuit construction developed in Section \ref{DQAOAFramework}.

\subsection{Monolithic QAOA}

Monolithic QAOA is a variational quantum algorithm introduced for approximate combinatorial optimization \cite{farhi2014quantum}. It prepares a parameterized quantum state by alternately applying a problem-dependent cost unitary and a mixing unitary, while a classical optimizer updates the circuit parameters to minimize the expected cost. In this mode, the full QUBO problem is implemented as one circuit on a single QPU. Thus, for a QUBO with $n$ binary variables, the QPU must have enough capacity to accommodate all $n$ qubits \cite{zhou2020quantum}.


Given the cost Hamiltonian $H_C$ (\ref{eq:cost_hamiltonian_pauli}), the standard QAOA ansatz of depth $p$ prepares a quantum state as
\begin{equation}
|\boldsymbol{\gamma},\boldsymbol{\beta}\rangle
=
U_M(\beta_p)U_C(\gamma_p)\cdots U_M(\beta_1)U_C(\gamma_1)|+\rangle^{\otimes n},
\label{eq:qaoa_state}
\end{equation}
where $\boldsymbol{\gamma}=(\gamma_1,\dots,\gamma_p)$ and $\boldsymbol{\beta}=(\beta_1,\dots,\beta_p)$ are the variational parameters of the ansatz, $U_C(\gamma)=e^{-i\gamma H_C}$ is the cost unitary as in \eqref{eq:qaoa_cost_unitary}, and $|+\rangle^{\otimes n}$ denotes the equal superposition over all computational basis states. In other words, the initial state assigns equal amplitude to all candidate binary solutions before the variational evolution is applied.

In the standard unconstrained formulation, the mixer Hamiltonian is chosen as
\begin{equation}
H_M=\sum_{i=1}^n X_i,
\label{eq:qaoa_mixer_hamiltonian}
\end{equation}
where $X_i$ denotes the Pauli-$X$ operator acting on qubit $i$. The Pauli-$X$ operator is the quantum analogue of a bit flip, mapping $|0\rangle$ to $|1\rangle$ and $|1\rangle$ to $|0\rangle$. This yields the mixer unitary
\begin{equation}
U_M(\beta)=e^{-i\beta H_M}.
\label{eq:qaoa_mixer_unitary}
\end{equation}
The role of the mixer is to redistribute amplitude among computational basis states, thereby enabling exploration of the binary search space \cite{farhi2014quantum,hadfield2019quantum}.

The QAOA parameters are determined by a classical optimizer that minimizes the expected value of the cost Hamiltonian over the variational state:
\begin{equation}
\min_{\boldsymbol{\gamma},\boldsymbol{\beta}}
\;
\langle \boldsymbol{\gamma},\boldsymbol{\beta}|H_C|\boldsymbol{\gamma},\boldsymbol{\beta}\rangle.
\label{eq:qaoa_outer_optimization}
\end{equation}
Here, $\langle \boldsymbol{\gamma},\boldsymbol{\beta}|H_C|\boldsymbol{\gamma},\boldsymbol{\beta}\rangle$ denotes the expectation value of the cost Hamiltonian with respect to the QAOA state. Equivalently, this quantity can be interpreted as the weighted average of the classical objective values over the measurement distribution induced by the variational state, i.e.,
\begin{equation}
\langle \boldsymbol{\gamma},\boldsymbol{\beta}|H_C|\boldsymbol{\gamma},\boldsymbol{\beta}\rangle
=
\sum_{\mathbf z\in\{0,1\}^n}
P_{\boldsymbol{\gamma},\boldsymbol{\beta}}(\mathbf z)\,F(\mathbf z),
\label{eq:qaoa_expectation_expanded}
\end{equation}
where $P_{\boldsymbol{\gamma},\boldsymbol{\beta}}(\mathbf z)$ is the probability of measuring bitstring $\mathbf z$ and $F(\mathbf z)$ is its corresponding QUBO objective value.

After optimization, the final state is measured in the computational basis, yielding bitstrings corresponding to candidate binary solutions. Increasing the circuit depth $p$ generally increases expressiveness, although it also increases circuit complexity and the cost of parameter optimization. In practice, QAOA is therefore studied as a depth-limited variational method that seeks a balance between solution quality and implementation cost on near-term quantum platforms \cite{farhi2014quantum,zhou2020quantum}.

Standard QAOA serves as the monolithic baseline for comparison with distributed mode. In this setting, all required cost and mixer operations are assumed to be implemented within a single QPU.



\subsection{Distributed Quantum Execution Concepts}

Distributed quantum execution refers to a setting in which a quantum algorithm is carried out across multiple interconnected QPUs rather than on a single QPU \cite{diadamo2021distributed}. In such architectures, qubits are partitioned across separate QPUs. Each QPU can perform local quantum operations on its own qubits, whereas interactions between qubits located on different QPUs require some form of distributed implementation or inter-QPU coordination \cite{barral2025review}.

For QUBO-based optimization, distributed execution becomes relevant once the binary variables are assigned to different QPUs. Under such an assignment, the pairwise couplings in the cost Hamiltonian can be divided into two categories: local couplings and cross-QPU couplings. A local coupling corresponds to a term whose two associated variables are placed on the same QPU, so the corresponding interaction can be executed locally. In contrast, a cross-QPU coupling corresponds to a term whose variables are placed on different QPUs, in which case the interaction cannot be implemented using only local gates \cite{sundaram2022distribution,mao2023qubit}.

One important way to realize such cross-QPU interactions is through the TeleGate protocol \cite{peckham2024asynchronous,sarvaghad2021general}. In this approach, the data qubits remain on their original QPUs, while an entangled communication pair shared across the two QPUs is used to implement the effect of a remote two-qubit gate. More specifically, the remote interaction is decomposed into local operations on each QPU together with measurements on the communication qubits and classical feedforward corrections determined by the measurement outcomes \cite{sarvaghad2021general}. In this way, the desired non-local gate is reproduced without physically moving the data qubits between QPUs. This is different from TeleData approaches, in which the state of a qubit is teleported so that the required gate can later be executed locally \cite{peckham2024asynchronous}. For distributed optimization circuits, the TeleGate viewpoint is especially relevant when the goal is to preserve variable-to-QPU placement while enabling interactions between variables residing on different QPUs.

This distinction makes the variable-to-QPU assignment an important structural element of distributed quantum computation. In particular, the interaction graph induced by the QUBO quadratic terms determines how many couplings remain local and how many become cross-QPU interactions. As a result, distributed execution is influenced not only by the problem size but also by how the problem structure is partitioned across QPUs. These concepts of local and cross-QPU interactions provide the basis for the DQAOA framework developed in the next section.

\section{Proposed DQAOA Framework and Simulator} \label{DQAOAFramework}
To solve a QUBO problem in a distributed QAOA setting, several connected blocks are required. First, the QUBO variables must be assigned to the available QPUs, because this assignment determines which interactions can be executed locally and which interactions require distributed implementation. Second, after the allocation is completed, the quadratic couplings of the QUBO problem must be classified as local or cross-QPU couplings. This classification is necessary because local couplings can be implemented with ordinary two-qubit gates, whereas cross-QPU couplings require remote operations and additional communication resources. Third, the corresponding QAOA circuit must be constructed and transpiled according to the selected execution mode, so that the same QUBO objective can be implemented either on a single QPU or across multiple QPUs. Fourth, the variational parameters of the circuit must be optimized using a consistent training objective, so that the monolithic and distributed modes remain directly comparable. Fifth, because QAOA performance can depend on circuit depth and initialization, the framework uses depth progression and multi-start search to improve parameter search robustness. Finally, after the optimized circuit is executed, the measured bitstrings must be decoded and evaluated to produce final solution and distribution-level metrics. These blocks form the complete workflow of the proposed simulator and are described in the following subsections.

\subsection{Multi-QPU Variable Allocation Strategies}

The first step in the distributed implementation is to assign QUBO variables to the available QPUs. The purpose of this allocation step is not to alter the underlying QUBO problem, but to determine how the same problem is embedded onto a distributed quantum architecture. Once an allocation is fixed using the process explained in this section, each quadratic coupling in the QUBO Hamiltonian inherits a hardware interpretation. If both variables of a coupling are assigned to the same QPU, the interaction is local. Otherwise, it becomes a cross-QPU interaction. For this reason, the quality of the variable allocation directly affects the communication burden of the distributed execution. 

Let the problem contain $n$ binary variables and let the distributed platform contain $M$ QPUs with capacities
\[
\mathbf{c} = [c_1,c_2,\dots,c_M],
\]
where $c_m$ denotes the maximum number of variable qubits that can be placed on QPU $m$. In the developed DQAOA simulator package, if the user does not provide the QPU capacities, they are generated automatically by \texttt{make\_balanced\_capacities}, allocating the capacities across available QPUs as evenly as possible based on the number of variables in the QUBO problem. Once the capacities are fixed, the allocation itself is built either by \texttt{make\_allocation}, which produces the default contiguous assignment, or by \texttt{make\_allocation\_from\_assignment}, which converts a user-specified QPU assignment into the internal allocation map used by the solver. The function \texttt{validate\_allocation} then checks that every variable is assigned exactly once, and no QPU capacity is exceeded, and that the required number of QPUs is actually used.

The final implementation supports three allocation options. The first is a contiguous allocation, in which variables are assigned to QPUs sequentially according to the available capacities. This provides a simple deterministic baseline and is constructed through \texttt{make\_allocation}. The second is a graph-aware allocation derived from the weighted interaction graph of the QUBO problem. In this case, \texttt{build\_weighted\_qubo\_adjacency} first constructs the weighted adjacency structure from the quadratic QUBO coefficients, and \texttt{graph\_aware\_greedy\_qpu\_assignment\_qubo} uses it to place strongly coupled variables on the same QPU whenever possible while still not violating capacity limits. The third strategy is manual allocation, in which the user directly specifies the QPU assignments of the variables, and the software converts them using \texttt{make\_allocation\_from\_assignment}. These alternatives allow the framework to study both straightforward and structure-aware allocation strategies within the same optimization workflow.

Rather than committing to one allocation immediately, the software first constructs candidate allocations using \texttt{build\_allocation\_candidates\_qubo}. For each candidate allocation, the software first counts the number of quadratic QUBO couplings connecting variables assigned to different QPUs. These are the cross-QPU terms, and they are important because each of them must later be realized through a distributed interaction using TeleGate rather than a local gate. The software also computes an estimated distributed interaction count at the selected circuit depth. In the current implementation, this estimate is
\[
2 \times (\text{number of cross-QPU quadratic terms}) \times p
\]
because each cross-QPU term is implemented through two remote controlled-NOT (CNOT) operations, and this pattern is repeated across all $p$ layers.

The candidate allocations are then ranked by \texttt{select\_best\_allocation\_candidate}. The primary criterion is the number of cross-QPU quadratic terms, so the preferred allocation is the one that minimizes the number of QUBO couplings that must be implemented across QPU boundaries. If candidate allocations remain tied after these counts are compared, the code applies a fixed priority order among the allocation types. In this way, the allocation step is guided by the amount of distributed communication the circuit will require.

\begin{figure}[H]
\centering
\begin{tikzpicture}[
    scale=0.95,
    every node/.style={transform shape},
    var/.style={circle, draw, thick, minimum size=7mm, inner sep=0pt, fill=white},
    qpubox/.style={draw, rounded corners, thick},
    weak/.style={black!70, line width=0.6pt},
    strong/.style={black, line width=1.2pt},
    localedge/.style={blue!70!black, line width=1.1pt},
    crossedge/.style={red!75!black, dashed, line width=1.1pt},
    paneltitle/.style={font=\small\bfseries},
    note/.style={font=\scriptsize},
    legendline1/.style={blue!70!black, line width=1.1pt},
    legendline2/.style={red!75!black, dashed, line width=1.1pt}
]

\begin{scope}[shift={(0,0)}]
\node[paneltitle] at (3.0,4.7) {(a) QUBO graph};

\node[var] (a1) at (1.0,3.2) {$z_1$};
\node[var] (a2) at (2.5,4.0) {$z_2$};
\node[var] (a3) at (4.0,3.2) {$z_3$};
\node[var] (a4) at (1.7,1.5) {$z_4$};
\node[var] (a5) at (3.3,1.5) {$z_5$};
\node[var] (a6) at (4.8,1.4) {$z_6$};

\draw[strong] (a1) -- (a4);
\draw[strong] (a4) -- (a5);
\draw[strong] (a1) -- (a5);

\draw[strong] (a2) -- (a3);
\draw[strong] (a3) -- (a6);
\draw[strong] (a2) -- (a6);

\draw[weak] (a2) -- (a5);
\draw[weak] (a1) -- (a2);

\node[note, align=center] at (3.0,0.45) {Two strongly coupled clusters are connected\\by a small number of weaker inter-cluster edges.};
\end{scope}

\begin{scope}[shift={(7.2,0)}]
\node[paneltitle] at (3.0,4.7) {(b) Contiguous allocation};

\draw[qpubox] (0.2,0.9) rectangle (2.8,4.0);
\draw[qpubox] (3.2,0.9) rectangle (5.8,4.0);

\node at (1.5,4.25) {\small QPU 1};
\node at (4.5,4.25) {\small QPU 2};

\node[var] (b1) at (1.0,3.5) {$z_1$};
\node[var] (b2) at (1.0,2.55) {$z_2$};
\node[var] (b3) at (1.0,1.6) {$z_3$};

\node[var] (b4) at (5.0,3.5) {$z_4$};
\node[var] (b5) at (5.0,2.55) {$z_5$};
\node[var] (b6) at (5.0,1.6) {$z_6$};

\draw[localedge] (b1) -- (b2);
\draw[localedge] (b2) -- (b3);
\draw[localedge] (b4) -- (b5);
\draw[localedge] (b5) -- (b6);

\draw[crossedge] (b1) -- (b4);
\draw[crossedge] (b1) -- (b5);
\draw[crossedge] (b2) -- (b5);
\draw[crossedge] (b2) -- (b6);

\node[note, align=center] at (3.0,0.45) {Cross-QPU terms = 4\\Estimated remote CX count = $8p$};
\end{scope}

\begin{scope}[shift={(0,-5.7)}]
\node[paneltitle] at (3.0,4.7) {(c) Graph-aware allocation};

\draw[qpubox] (0.2,0.9) rectangle (2.8,4.0);
\draw[qpubox] (3.2,0.9) rectangle (5.8,4.0);

\node at (1.5,4.25) {\small QPU 1};
\node at (4.5,4.25) {\small QPU 2};

\node[var] (c1) at (1.0,3.5) {$z_1$};
\node[var] (c4) at (1.0,2.55) {$z_4$};
\node[var] (c5) at (1.0,1.6) {$z_5$};

\node[var] (c2) at (5.0,3.5) {$z_2$};
\node[var] (c3) at (5.0,2.55) {$z_3$};
\node[var] (c6) at (5.0,1.6) {$z_6$};

\draw[localedge] (c1) -- (c4);
\draw[localedge] (c4) -- (c5);
\draw[localedge] (c1) to[bend right=45] (c5);

\draw[localedge] (c2) -- (c3);
\draw[localedge] (c3) -- (c6);
\draw[localedge] (c2) to[bend left=45] (c6);

\draw[crossedge] (c2) -- (c5);
\draw[crossedge] (c1) -- (c2);

\node[note, align=center] at (3.0,0.45) {Cross-QPU terms = 2\\Estimated remote CX count = $4p$};
\end{scope}

\begin{scope}[shift={(7.2,-5.7)}]
\node[paneltitle] at (3.0,4.7) {(d) Manual allocation};

\draw[qpubox] (0.2,0.9) rectangle (2.8,4.0);
\draw[qpubox] (3.2,0.9) rectangle (5.8,4.0);

\node at (1.5,4.25) {\small QPU 1};
\node at (4.5,4.25) {\small QPU 2};

\node[var] (d1) at (1.0,3.5) {$z_1$};
\node[var] (d2) at (1.0,2.55) {$z_2$};
\node[var] (d5) at (1.0,1.6) {$z_5$};

\node[var] (d3) at (5.0,3.5) {$z_3$};
\node[var] (d4) at (5.0,2.55) {$z_4$};
\node[var] (d6) at (5.0,1.6) {$z_6$};

\draw[localedge] (d1) -- (d2);
\draw[localedge] (d1) to[bend right=45] (d5);
\draw[localedge] (d2) -- (d5);
\draw[localedge] (d3) to[bend left=45] (d6);

\draw[crossedge] (d1) -- (d4);
\draw[crossedge] (d4) -- (d5);
\draw[crossedge] (d2) -- (d3);
\draw[crossedge] (d2) -- (d6);

\node[note, align=center] at (3.0,0.45) {Cross-QPU terms = 4\\Estimated remote CX count = $8p$};
\end{scope}

\draw[legendline1] (2.0,-6.5) -- (2.8,-6.5);
\node[note, anchor=west] at (2.95,-6.5) {local coupling};

\draw[legendline2] (6.1,-6.5) -- (6.9,-6.5);
\node[note, anchor=west] at (7.05,-6.5) {cross-QPU coupling};

\end{tikzpicture}
\caption{Illustration of the variable allocation strategies used in the DQAOA framework on an example six-variable QUBO graph. Panel (a) shows the original interaction graph, which contains two strongly coupled clusters connected by a small number of weaker inter-cluster edges. Panels (b)--(d) show the resulting variable-to-QPU assignments under contiguous, graph-aware, and manual allocation, respectively. Solid blue edges correspond to local couplings, while dashed red edges correspond to cross-QPU couplings. The graph-aware strategy reduces cross-QPU interactions by placing strongly coupled variables on the same QPU whenever possible.}
\label{fig:allocation_strategies}
\end{figure}
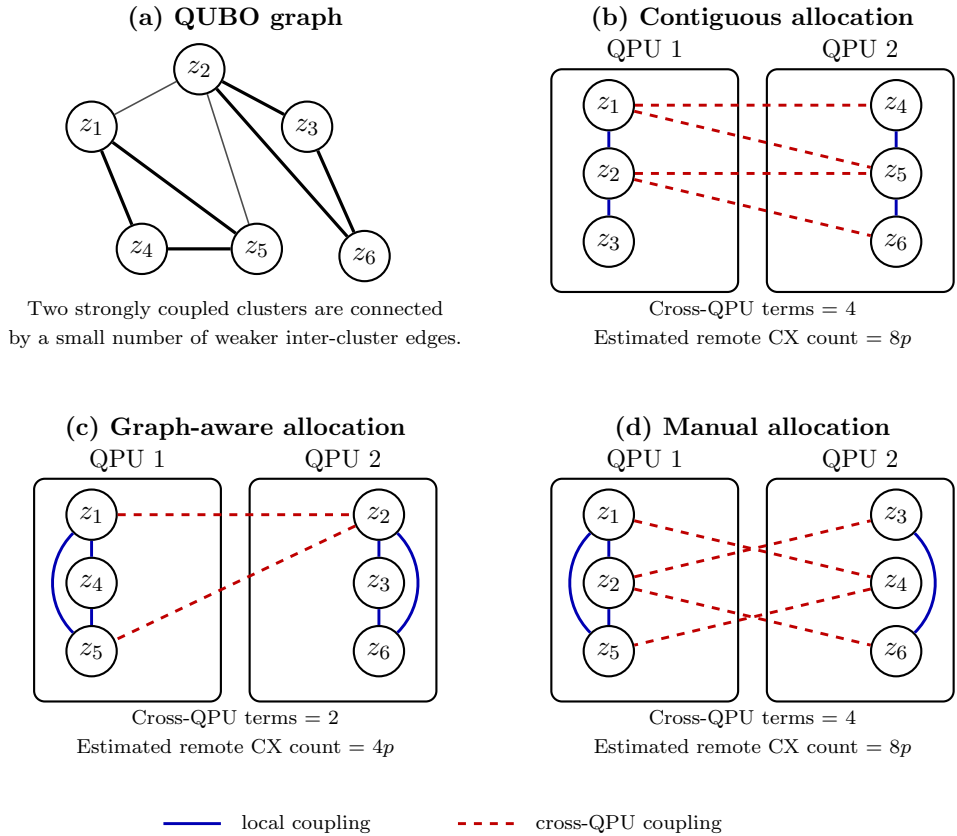

This variable-to-QPU assignment serves as the basis for all subsequent distributed processing. Once the allocation is fixed, the framework can identify local and cross-QPU couplings explicitly and then construct the corresponding distributed circuit under the selected execution mode. Fig.~\ref{fig:allocation_strategies} illustrates these allocation strategies on a six-variable QUBO graph and shows how the resulting variable placement affects the number of cross-QPU couplings.

\subsection{Identifying Local and Cross-QPU Couplings}

Once a variable to QPU allocation is determined, each quadratic coupling in the canonical QUBO form is classified according to the QPUs of its two associated variables. Recall that \(b_{ij}\) denotes the coefficient (weight) of the quadratic term \(z_i z_j\) in the canonical QUBO representation for \(i<j\). Let \(q(i)\) denote the QPU to which variable \(z_i\) is assigned. Then, for each quadratic term \(b_{ij} z_i z_j\), the corresponding coupling is identified as
\begin{equation}
\text{local if } q(i)=q(j)\hspace{15pt}, \qquad
\text{cross-QPU if } q(i)\neq q(j)
\label{eq:local_cross_rule}
\end{equation}

In the final implementation, this classification is used in two ways. First, the function \texttt{count\_cross\_qpu\_quadratic\_terms} counts the number of quadratic QUBO terms that connect variables assigned to different QPUs. This quantity is used during allocation evaluation and is also reported later as part of the distributed execution statistics. Second, the same classification is applied inside the circuit construction functions \texttt{build\_qubo\_qaoa\_circuit} and \texttt{build\_parametrized\_qubo\_qaoa\_circuit}. For each quadratic term, the code checks whether the two variables belong to the same QPU. If they do, the corresponding interaction is implemented through the local \(ZZ\) decomposition. Otherwise, it is treated as a cross-QPU interaction and is passed to the distributed remote-gate realization.

This local versus cross-QPU decision is therefore the implementation step that connects the allocation stage to the later distributed circuit. Once the allocation is fixed, the set of quadratic QUBO terms is partitioned into interactions that can be executed locally and interactions that must be implemented through TeleGate-based remote operations across QPU boundaries. This distinction is also illustrated in Fig.~\ref{fig:allocation_strategies}, where solid blue edges correspond to local couplings and dashed red edges correspond to cross-QPU couplings under each allocation. The classification of local and cross-QPU couplings serves as the direct input to the distributed circuit construction stage described next.

\subsection{Distributed Circuit Construction and Transpilation}

Once the allocation is done and the quadratic QUBO terms have been classified as local or cross-QPU, the next step is to construct the corresponding quantum circuit. To carry out this task, we have developed two functions \texttt{build\_qubo\_qaoa\_circuit} and \texttt{build\_parametrized\_qubo\_qaoa\_circuit}. Both functions take as input the canonical QUBO, the circuit depth \(p\), and the selected allocation approach. Their role is to map the canonical QUBO objective into the layered QAOA circuit. The first function constructs a circuit with the numerical values of the QAOA angles directly inserted into the circuit, whereas the second constructs a circuit with symbolic parameter vectors for the variational parameters (\(\gamma\) and \(\beta\)), so that the circuit structure can later be reused during repeated objective evaluations. 

The circuit is constructed from the canonical QUBO instance by assigning one data qubit to each binary decision variable and then translating the linear and quadratic QUBO coefficients into the corresponding QAOA cost-layer operations. In the distributed mode, the circuit also includes two communication qubits and a classical register. The communication qubits are used to implement remote two-qubit operations, such as CNOT gates, between data qubits located on different QPUs. The classical register stores both the final measurement outcomes of the data qubits and the intermediate measurement outcomes needed for the conditional corrections applied during remote gate execution. In the final implementation, the number of additional classical bits is determined from the total number of remote COT operations required in the circuit, which depends on the number of cross-QPU quadratic terms and the circuit depth. 

After the registers are created, the data qubits are initialized in the equal superposition state by applying a Hadamard gate to every data qubit, and the QAOA layers are then assembled one by one. For each layer, the implementation first inserts the cost-unitary terms associated with the linear and quadratic coefficients of the canonical QUBO objective and then applies the mixer layer. Each linear coefficient contributes a single-qubit \(R_Z\) rotation on the corresponding data qubit, while each quadratic coefficient contributes the \(ZZ\) interaction associated with the corresponding pair of variables. If the two variables belong to the same QPU, this interaction is implemented locally using the usual sequence of CNOT, a phase rotation, and another CNOT. After all cost terms of the layer are inserted, the mixer is applied through \(R_X(2\beta)\) rotations on all data qubits. The purpose of this step is to redistribute amplitude among computational-basis states, so that the circuit can explore different candidate bitstrings rather than only accumulate cost-dependent phase information. In this way, the mixer works together with the cost unitary to guide the measurement distribution toward lower-cost solutions. This procedure is repeated for all \(p\) layers of the quantum circuit. 

The distributed part of the circuit appears when a quadratic QUBO term connects variables assigned to different QPUs. In that case, the local two-qubit interaction used in the \(ZZ\) decomposition cannot be applied directly and is replaced by TeleGate-based remote gate implementation by \texttt{apply\_remote\_cx\_telegate\_explicit}. To illustrate how this process works, consider a control data qubit on $QPU_{A}$ and a target data qubit on $QPU_{B}$. The procedure first resets the two communication qubits, prepares one of them in superposition, and entangles the communication pair. One communication qubit is then coupled locally to the control data qubit on $QPU_{A}$ and measured, while the other communication qubit is coupled locally to the target data qubit $QPU_{B}$ and measured as well. The two measurement outcomes are stored in classical bits and are then used to trigger conditional Pauli corrections on the data qubits. In this way, the effect of a remote CNOT is reproduced without moving the logical data qubits away from their assigned QPUs. Since each cross-QPU \(ZZ\) interaction requires two such remote CNOT operations, every cross-QPU quadratic term introduces additional circuit depth, communication qubit usage, measurements, reset operations, and classical feedforward. 

\begin{figure}[H]
  \centering

  \begin{subfigure}{\linewidth}
    \centering
    \includegraphics[width=\linewidth]{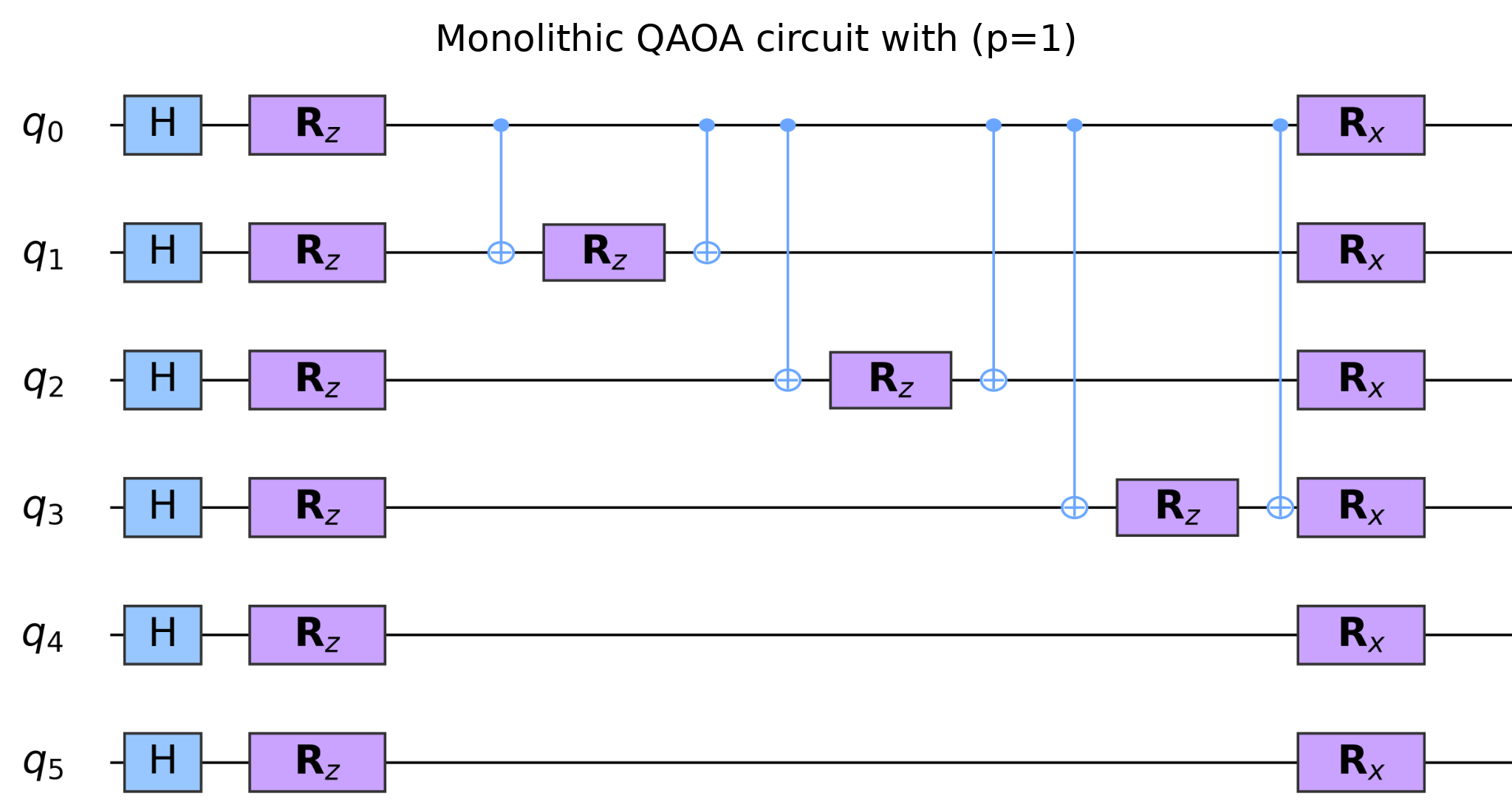}
    \caption{Monolithic QAOA ansatz on a single QPU with local implementation of all linear and quadratic cost terms.}
    \label{fig:mono-qaoa-ansatz}
  \end{subfigure}

  \vspace{1.5em}

  \begin{subfigure}{\linewidth}
    \centering
    \includegraphics[width=\linewidth]{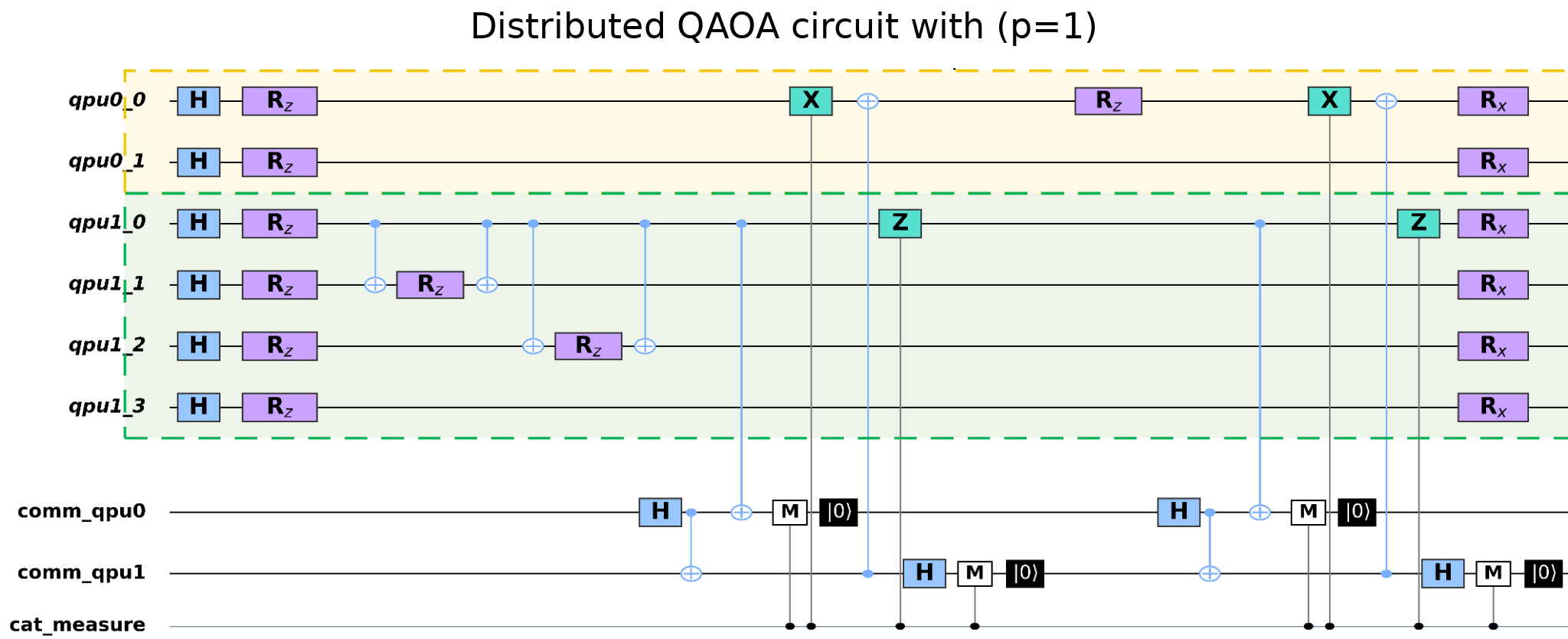}
    \caption{Distributed QAOA ansatz across two QPUs with communication qubits, measurements, and conditional corrections for cross-QPU entanglements. Yellow region indicates the first QPU while green area indicates the second QPU.}
    \label{fig:dqaoa-ansatz}
  \end{subfigure}

  \caption{Comparison of monolithic and distributed QAOA circuit structures.}
  \label{fig:qaoa-ansatz-comparison}
\end{figure}

We do not rebuild and retranspile the circuit from scratch during every objective evaluation. Instead, the parameterized circuit is passed to \texttt{CachedQuboCircuitRunner}, which performs one-time circuit construction and one-time transpilation for a fixed problem instance, depth, allocation, distributed mode, and backend configuration. The helper function \texttt{get\_cached\_qubo\_runner} manages this reuse through a cache key based on the structural QUBO signature and the relevant execution settings. Once the cached runner is available, \texttt{bind\_parameters} inserts the current QAOA angles into the already transpiled circuit, while \texttt{run\_counts} and \texttt{run\_counts\_batch} execute one or several bound circuits under the same circuit structure. As a result, circuit construction and transpilation become setup steps associated with a fixed circuit structure rather than repeated work inside the variational loop. This is the main implementation mechanism that links the circuit stage to the runtime reductions reported later in the paper. Fig.~\ref{fig:qaoa-ansatz-comparison} illustrates how the framework constructs the distributed QAOA circuit, distinguishes local and cross-QPU interactions, and then reuses the transpiled parameterized circuit during repeated evaluations.

\subsection{Training Objective and Variational Parameter Optimization}

In both monolithic and distributed modes, the variational parameters are optimized with respect to the same training objective, ensuring that all execution settings remain directly comparable. At circuit depth $p$, let
\[
\mathbf{x}=[\gamma_1,\beta_1,\gamma_2,\beta_2,\dots,\gamma_p,\beta_p]
\]
denote the vector of QAOA parameters. For a given parameter vector, the corresponding circuit is executed and measured, which produces a probability distribution over bitstrings. The training objective is then defined as the expected value of the QUBO cost under that measurement distribution.

Using the canonical QUBO objective function $F(\mathbf z)$ derived in \eqref{eq:qubo_canonical_work}, the training objective can be written as
\begin{equation}
J(\mathbf{x})
=
\sum_{\mathbf z\in\{0,1\}^n}
P_{\mathbf{x}}(\mathbf z)\,F(\mathbf z),
\label{eq:training_objective_mean_cost}
\end{equation}
where $P_{\mathbf{x}}(\mathbf z)$ denotes the probability of measuring bitstring $\mathbf z$ when the circuit is executed with parameter vector $\mathbf{x}$. In the implementation, this quantity is estimated from shot-based measurement counts, and the same sampled mean-cost objective is used consistently in all quantum modes. Therefore, the parameter optimization problem is
\begin{equation}
\mathbf{x}^{\star}
=
\arg\min_{\mathbf{x}} J(\mathbf{x}).
\label{eq:training_objective_argmin}
\end{equation}
where $\mathbf{x^{*}}$ is the vector of optimal variational parameters. 

In the final implementation, the objective evaluator is constructed for a fixed problem instance, circuit depth, allocation, and execution mode, and is then reused throughout the corresponding optimization run. For each parameter vector, the evaluator executes the circuit, converts the resulting counts into a probability distribution over bitstrings, and computes the sampled mean cost. When multiple averaging passes are requested, the same evaluation is repeated over independent shot-based executions, and the resulting mean-cost values are averaged. The purpose of this averaging is to reduce the variance introduced by finite-shot sampling, thereby making the estimated training objective more stable and reliable for variational parameter updates. In addition to the training objective itself, the evaluator also records diagnostic quantities, such as the best observed sampled cost and the probability mass concentrated on low-cost solutions. These quantities are not used directly in the parameter update, but they provide useful information about how the sampled distribution changes during the search.

The variational parameters are updated using an Adam optimizer with a Simultaneous Perturbation Stochastic Approximation (SPSA) gradient estimator. At each iteration, two perturbed parameter vectors are generated around the current point, and the difference between the corresponding objective values is used to form a stochastic gradient estimate with respect to all parameters simultaneously. This gradient estimate is then used in the Adam update rule to obtain the next parameter vector. In the present implementation, the objective at the current parameter vector can also be evaluated within the same iteration so that the optimization history records the actual sampled objective value along the search trajectory. Compared with coordinate-wise finite-difference estimation, this strategy requires substantially fewer objective evaluations per iteration and is therefore better suited to shot-based variational optimization.

Because the same expected cost objective and the same parameter update rule are used in both the monolithic and distributed modes, the resulting comparisons remain consistent across execution settings. The manner in which this parameter optimization is carried out across circuit depths and multiple initializations is described in the next subsection.

\subsection{Depth Progression and Multi-Start Search}

The proposed framework does not optimize the quantum circuit at only one fixed depth. Instead, the search is carried out progressively from the lowest depth, $p=1$, up to a predefined maximum depth $p_{\max}$. This procedure is implemented inside \texttt{run\_single\_qaoa\_mode}, which loops over the depth values and performs a separate parameter search at each depth. For every depth, the function first constructs the corresponding objective evaluator using \texttt{build\_objective\_fn\_qubo} and then executes the parameter-search process for that depth. In this way, the optimization starts from shallower circuits and gradually moves toward deeper, more expressive ones.

To transfer information across depths, the best parameter vector obtained at depth $p-1$ is first expanded to the parameter dimension of depth $p$ by \texttt{lift\_angles\_to\_next\_depth}. The function \texttt{generate\_initial\_points\_for\_depth} then uses this transferred parameter vector to construct the initial points for the next depth. More specifically, the transferred vector itself can be included directly as a warm-start initialization, and additional warm-start initializations are produced by applying small random perturbations around it. The same function also generates independent random initializations at the current depth, and the user can specify the number of such random starts. As a result, the search at each depth combines information carried forward from the previous depth with fresh exploration from new random starting points.

Each initialization yields an independent optimization run. In the final implementation, these runs are executed either sequentially or in parallel, depending on the value of \texttt{parallel\_restarts} in \texttt{MultiStartConfig}. After all runs at the current depth are completed, their outputs are collected in a candidate set. The ranking and selection of these candidates are implemented by \texttt{candidate\_rank\_tuple} and \texttt{choose\_best\_result}. When a classical reference is available and the setting \texttt{select\_by\_exact\_when\_available} is enabled, the ranking first prioritizes agreement with the exact optimum solution. In that case, the candidates are ordered first by whether the sampled best cost matches the exact optimum cost provided by the classical reference, then by whether the sampled best bitstring matches the exact optimum bitstring, then by the probability mass assigned to the exact optimum cost, and finally by the probability assigned to the exact optimum bitstring. If candidates remain tied after these criteria are checked, the ranking next compares their sampled quantum performance. It first considers the best observed sampled cost, then the probability mass assigned to that cost, then the probability of the corresponding best sampled bitstring, followed by the sampled mean cost.

When a classical reference is unavailable, the ranking is based solely on sampled quantum performance. In that case, the primary criterion is the best observed sampled cost. If multiple candidates attain the same best sampled cost, preference is then given to the candidate with a larger probability mass assigned to that cost, followed by a larger probability of the associated best sampled bitstring. If a tie remains, the ranking then prefers the smaller sampled mean cost and distribution gap. Therefore, even without a classical reference, the framework still applies a well-defined ranking rule to compare candidate runs. After the candidate runs at a given depth are ranked, \texttt{choose\_best\_result} selects the best-performing one as the representative for that depth. The selected parameter vector is retained and used to warm-start the search at the next depth. After all depths from $1$ to $p_{\max}$ have been processed, the retained depth representatives are compared again using the same ranking logic, and the best among them is reported as the final result for the selected execution mode.

This combination of depth progression, warm-start initialization, and random restarts is used because the QAOA optimization landscape is generally non-convex, and the final result can depend strongly on the starting point. The warm-start initializations help preserve useful information obtained at lower depths, whereas the random starts reduce the risk that the search remains confined to a poor local solution. Since the same depth progression and multi-start procedure are used in both monolithic and distributed modes, the resulting comparisons remain consistent across execution settings. Algorithm~\ref{alg:depth_multistart_dqaoa} summarizes the depth progression and multi-start search procedure implemented in the final DQAOA framework.

\subsection{Measurement Decoding and Final Distribution Metrics}
After each optimization run is completed, the final parameter vector is evaluated using a separate shot-based circuit execution with the selected mode and the predefined number of final shots, set by the user. The purpose of this step is to assess the optimized circuit in a single, consistent final evaluation run, rather than relying solely on the noisier objective estimates encountered during training. In the implementation, this step is performed within \texttt{evaluate\_qaoa\_candidate}, which executes the final circuit and computes the final sampled metrics for that run.

In the monolithic QAOA mode, the measurement counts returned by the final circuit execution are not yet in the logical QUBO bitstring order. The data qubits are measured directly into the classical register, but under Qiskit's little-endian convention, the displayed measurement bitstring is reversed relative to the qubit-index order. For example, if the displayed measurement bitstring is written as $(b_{n-1}b_{n-2}\dots b_1 b_0)$, then the corresponding logical QUBO bitstring is recovered as $(b_0 b_1 \dots b_{n-1})$. Therefore, the implementation reverses the displayed bitstring to recover the logical QUBO bitstring.

In the distributed QAOA mode, the decoding step is more involved because the classical register contains not only the measurement outcomes of the data qubits, but also additional bits associated with the communication qubits and the conditional correction steps used for cross-QPU interactions. For this reason, the implementation first removes any formatting spaces from the displayed measurement bitstring, then reverses the full measured classical bitstring to restore the logical bit order, and finally keeps only the first $n$ bits, which correspond to the data qubits of the QUBO problem. This decoding step is performed by \texttt{bitstring\_from\_counts\_key} and \texttt{extract\_data\_bits\_from\_full\_count\_key}.

After this step, the functions of \texttt{decoded\_distribution\_from\_counts} and \texttt{decoded\_count\_map\_from\_counts} combine all measured outcomes that correspond to the same logical QUBO bitstring and produce the final probability distribution and count map over logical QUBO bitstrings. Once the decoded distribution is available, the framework evaluates the QUBO objective on the observed samples. The function \texttt{counts\_qubo\_stats} computes the best observed sampled bitstring, the best observed sampled cost, the sampled mean cost, and the probability of the best sampled bitstring. In addition, \texttt{low\_cost\_mass\_from\_distribution} computes the probability mass assigned to samples whose cost lies within a margin of the best observed sampled cost. To provide a more detailed view of the final distribution, \texttt{build\_elite\_metrics\_from\_counts} also constructs the ordered set of top sampled bitstrings together with their costs, counts, and frequencies. These quantities are then collected by \texttt{compute\_final\_distribution\_metrics}, which forms the final package of sampled performance metrics for the run.

When a classical reference is provided, the same decoded distribution can also be evaluated relative to that reference. In that case, the implementation computes quantities such as the probability mass assigned to the exact optimum cost, the probability assigned to the exact optimum bitstring, and the agreement of the best sampled quantum results with the classical optimum results. These reference-based quantities are combined with the sampled metrics using \texttt{build\_convergence\_report} to form the final summary for the run. The output of this stage is therefore not only a single best-sampled bitstring, but also a structured set of distribution-level metrics derived from the final shot-based evaluation. These metrics are the quantities the framework subsequently uses to compare candidate runs and report final results.

\begin{algorithm}[H]
\caption{Depth progression and multi-start search in the DQAOA framework}
\label{alg:depth_multistart_dqaoa}
\begin{algorithmic}[1]
\Require QUBO instance, execution mode, maximum depth $p_{\max}$, optimization settings, and optional exact classical reference
\Ensure Final selected parameter vector and associated sampled solution metrics

\State Set previous-depth best parameter vector to None
\State Initialize retained depth representatives $\mathcal{R} \gets \varnothing$

\For{$p = 1$ to $p_{\max}$}
    \State Construct the objective evaluator for depth $p$ 
    \State Generate the set of initial points for depth $p$
    \If{a previous-depth best parameter vector exists}
        \State Lift it to depth $p$ and include warm-start initializations
        \State Add perturbed warm-start initializations
    \EndIf
    \State Add the requested number of random initializations

    \State Initialize candidate set $\mathcal{C}_p \gets \varnothing$
    \For{each initial point in the generated set}
        \State Run the variational optimization from that initial point
        \State Perform the final shot-based evaluation of the optimized circuit
        \State Store the resulting candidate in $\mathcal{C}_p$
    \EndFor

    \State Rank candidates in $\mathcal{C}_p$
    \State Select the best candidate at depth $p$
    \State Store the selected candidate in $\mathcal{R}$
    \State Retain its parameter vector as the previous-depth best parameter vector
\EndFor

\State Rank the retained depth representatives in $\mathcal{R}$ using the same rule
\State Select the best retained representative as the final output
\State \Return final selected parameter vector and associated sampled solution metrics
\end{algorithmic}
\end{algorithm}

\subsection{General Solver Modes and Package Workflow}

Beyond the individual algorithmic components described above, this work's final contribution is a general DQOA framework and its simulator package, enabling users to solve QUBO problems across multiple execution settings within a single, consistent workflow. The package accepts a QUBO instance in dense form via the quadratic matrix, linear vector, and constant term, converts it to the canonical internal representation, and then applies the selected solver mode. In this way, the package serves not only as a distributed QAOA simulator, but also as a common experimental platform for solving and comparing QUBO problems under classical, monolithic QAOA, and distributed QAOA settings.

At the user level, the main solver interfaces are \texttt{solve\_qubo\_mode} and \texttt{compare\_qubo\_modes}. The package supports brute-force search, CPLEX-based mixed-integer optimization, monolithic QAOA, and distributed QAOA within a single interface. As a result, a user can solve the same QUBO instance with monolithic QAOA and distributed QAOA modes, compare their final sampled solutions and costs, and examine how the execution setting affects runtime, cross-QPU interaction count, and distribution-level solution quality. This unified interface is one of the main practical contributions of the package, since it removes the need to implement separate workflows for each execution mode.

The package also exposes several user-configurable parameter groups that control different stages of the workflow. The training settings in \texttt{TrainConfig} determine quantities such as the number of training shots, the number of averaging passes, the simulator and transpiler settings, and whether batch evaluation is used. Increasing the number of training shots or averaging passes generally makes the sampled objective estimate more stable, but it also increases runtime. The optimization settings in \texttt{AdamConfig} control the number of iterations, the learning rate, and the SPSA perturbation parameters. Increasing the number of iterations may improve the final parameter search, but again at the cost of longer runtime. The multi-start settings in \texttt{MultiStartConfig} control the number of random initializations, the number and size of warm-start perturbations, whether the plain warm-start point is included, and how many restart runs are executed in parallel. A larger number of multi-starts can improve robustness against poor local solutions, but they require more optimization runs, which increases runtime. \texttt{AnalysisConfig} controls how the final sampled results are reported, including the number of best sampled bitstrings retained in the final summary and the tolerance used in reference-based checks. \texttt{ConvergenceConfig} controls how the optimization history is assessed for stabilization. These settings do not change the optimization problem itself, but only how the run is analyzed and summarized.

From a workflow perspective, the package first canonicalizes the input QUBO, then selects the requested solver mode, constructs the corresponding circuit or classical solver path, performs the variational search when a QAOA mode is used, evaluates the final sampled distribution, and finally returns structured outputs summarizing the solution quality. In addition to the core solver functions, the package provides visualization tools, including tabulated mode comparisons, detailed result summaries, elite-bitstring plots, and circuit-figure generation. These tools make it easier for users to inspect not only the final best solution, but also the distribution-level behavior of each mode. Together with the runtime optimization strategies discussed in the next section, the package facilitates more efficient solutions and comparisons of QUBO problems within a general software framework.

To improve practical accessibility, the final software package also includes a Streamlit-based graphical user interface, referred to as the DQAOA-QUBO Solver Studio. This interface wraps the same backend functions used by the Python package, but allows the user to interact with the solver through a web-based dashboard. Users can select built-in QUBO examples or provide custom QUBO inputs through an editable matrix, pasted text, or CSV upload. The sidebar exposes the main configuration settings, including the number of variables, solver modes, QAOA depth, training and final sampling shots, optimizer settings, multi-start settings, number of QPUs, and QPU capacities. The interface also validates the selected QPU configuration before execution, ensuring that monolithic QAOA uses a single QPU and distributed QAOA uses multiple QPUs with sufficient total capacity. After execution, the dashboard reports the mode comparison table, best bitstring, best cost, solution accuracy plots, elite sampled bitstrings, and final optimized variational parameters. This graphical layer does not change the underlying solver logic, but makes the developed simulator easier to use, reproduce, and demonstrate.

\begin{figure}[H]
\centering
\usetikzlibrary{positioning, arrows.meta, calc}
\resizebox{0.8\linewidth}{!}{%
\begin{tikzpicture}[
    node distance=0.55cm and 0.55cm,
    box/.style={draw, rounded corners, align=center, minimum height=0.9cm, minimum width=3.5cm},
    smallbox/.style={draw, rounded corners, align=center, minimum height=0.75cm, minimum width=1.45cm},
    arrow/.style={-Latex, thick},
    every node/.style={font=\scriptsize}
]

\node[box] (input) {QUBO input\\$(H,f,c_0)$};

\node[box, below=of input] (config) {User settings\\training, optimizer, multi-start settings};

\node[box, below=of config] (interface) {Main package interface\\\texttt{compare\_qubo\_modes} / \texttt{solve\_qubo\_mode}};

\node[smallbox, below left=0.7cm and 0.9cm of interface] (bf) {Brute\\force};
\node[smallbox, below right=0.7cm and 0.9cm of interface] (cplex) {CPLEX\\MIQP};
\node[smallbox, below=0.7cm of interface, xshift=-1.75cm] (mono) {Monolithic\\QAOA};
\node[smallbox, below=0.7cm of interface, xshift=1.75cm] (dist) {Distributed\\QAOA};

\node[box, below=0.75cm of mono, xshift=1.75cm] (setup) {Allocation and local/cross-QPU coupling identification};

\node[box, below=of setup] (circuit) {Circuit construction and transpilation};

\node[box, below=of circuit] (train) {For QAOA modes:\\training and search\\objective evaluation, SPSA-Adam, depth progression, multi-start\\(Algorithm~\ref{alg:depth_multistart_dqaoa})};

\node[box, below= of train] (eval) {Final evaluation and comparison\\bitstring decoding, sampled metrics, reference-based checks};

\node[box, below=of eval] (out) {Outputs\\best bitstring, best cost, summaries, tables, plots};

\draw[arrow] (input) -- (config);
\draw[arrow] (config) -- (interface);

\draw[arrow] (interface.west) -- ++(-1.65,0) -| (bf.north);
\draw[arrow] (interface.east) -- ++(1.65,0) -|(cplex.north);
\draw[arrow] (interface.south) ++ (-1.75,0) -- (mono.north);
\draw[arrow] (interface.south) ++ (1.75,0) -- (dist.north);

\draw[arrow] (mono.south) -- ($(setup.north west)!(mono.south)!(setup.north east)$);
\draw[arrow] (dist.south) -- ($(setup.north west)!(dist.south)!(setup.north east)$);
\draw[arrow] (setup) -- (circuit);
\draw[arrow] (circuit) -- (train);
\draw[arrow] (train) -- (eval);

\draw[arrow] (bf.west) -- ++(-0.7,0) |- (eval.west);
\draw[arrow] (cplex.east) -- ++(0.7,0) |- (eval.east);

\draw[arrow] (eval) -- (out);

\end{tikzpicture}%
}
\caption{High-level workflow of the DQAOA software package. Users provide QUBO instances and configuration settings, select one or more solver modes, and obtain final solution summaries and comparison outputs within one consistent workflow. For QAOA modes, the package also performs allocation, circuit construction and transpilation, variational parameter optimization, and final distribution-level evaluation.}
\label{fig:package_workflow}
\end{figure}
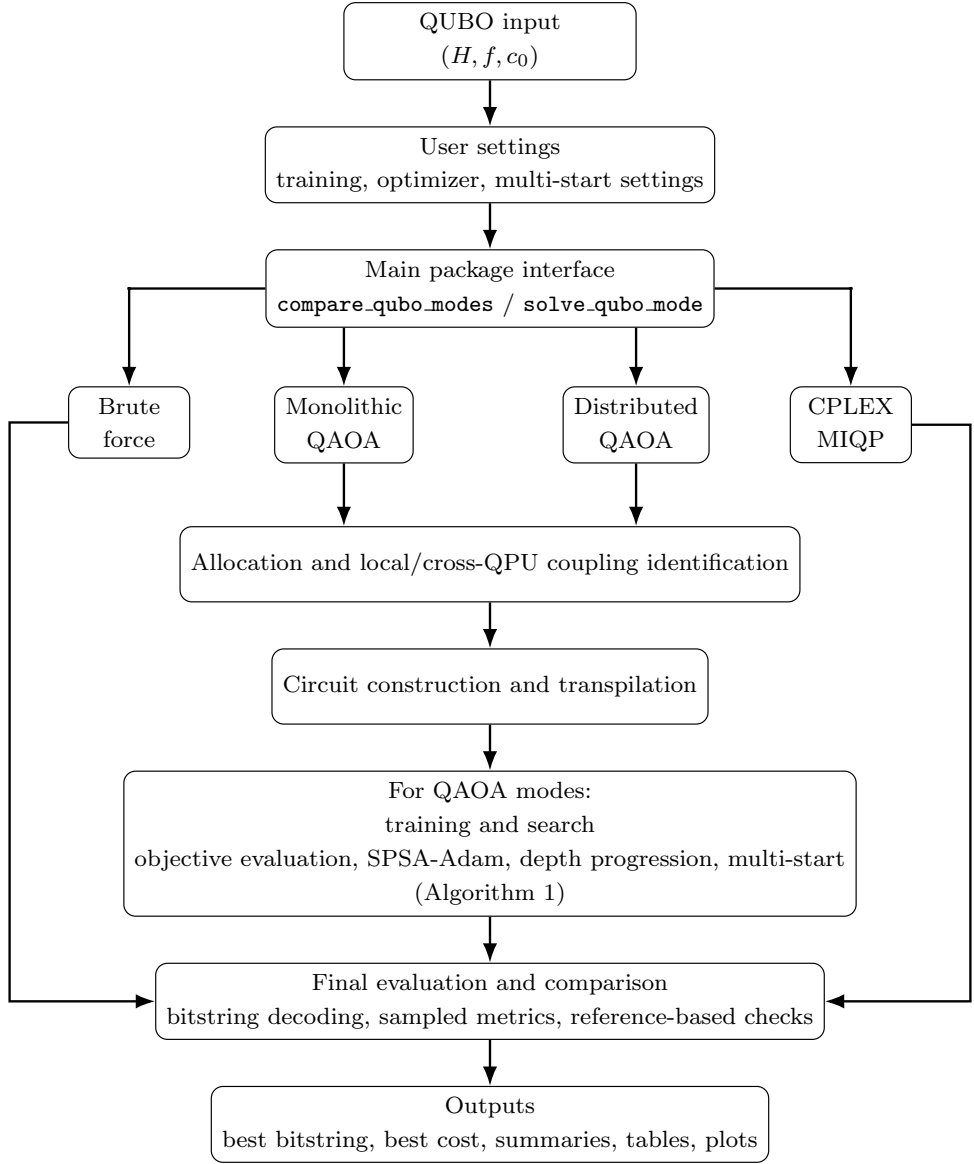

The software is open source and publicly available through the \href{https://sites.google.com/site/aminkargarian/home}{RAISE LAB website} and \href{https://github.com/LSU-RAISE-LAB/}{RAISE LAB GitHub repository}. Users can access the package there and consult the accompanying README documentation for installation instructions, required inputs, configurable parameters, and example workflows. The motivation for developing this DQAOA simulator package was the lack of a publicly available, reusable software workflow that could handle general QUBO instances consistently across both monolithic and distributed QAOA settings. In this sense, the package acts as an enabling research tool. It does not change the underlying QUBO problem or the QAOA objective itself, but it makes distributed quantum optimization studies more accessible, reproducible, and practically testable across different solver modes. The overall workflow of the package is summarized in Fig.~\ref{fig:package_workflow}.

\section{Runtime Optimization Strategies}

The initial implementation of the QUBO solver was functionally correct, but its runtime was dominated by repeated work inside the variational loop. In particular, the main sources of overhead were repeated circuit construction and transpilation, a large number of objective evaluations during parameter updates, repeated reconstruction of objects that did not change across optimization runs, and limited throughput across repeated starts. To address these issues, the final implementation introduces a set of runtime-oriented improvements that preserve the QUBO objective, execution modes, and final evaluation logic while reducing the workflow computational cost.

\subsection{Reducing Circuit Evaluation Overhead}

The first optimization targets the most expensive repeated task in the variational loop, which is rebuilding and transpiling quantum circuits for every objective evaluation. In the final implementation, the circuit for a given problem instance, depth, allocation, and execution mode is constructed once as a parameterized circuit and transpiled once for the selected backend (either a simulator or hardware). During optimization, circuit evaluations are performed by binding updated parameter values to the transpiled circuit rather than reconstructing the full circuit from scratch. This change removes substantial avoidable overhead from the inner loop and makes repeated objective evaluations substantially cheaper.

To enable reliable reuse of previously compiled circuits, the final implementation checks whether a newly requested solver run corresponds to a problem and execution setting that have already been processed. This check is based on the QUBO coefficients and the relevant execution settings, including circuit depth, variable allocation, execution mode, and execution platform configuration. When these quantities remain unchanged, the implementation reuses the previously prepared circuit rather than rebuilding and retranspiling it. This reduces unnecessary repeated work when the same QUBO problem is evaluated multiple times under the same settings.

\subsection{Reducing Optimization Overhead}

A second group of improvements reduces the number of expensive objective evaluations required during variational parameter optimization. In the original finite-difference Adam implementation, the gradient with respect to the variational parameters was estimated by perturbing one parameter at a time and reevaluating the objective for each perturbation. As the number of circuit parameters increased with depth, this procedure required a growing number of separate objective evaluations per iteration, thereby becoming increasingly expensive. In the final implementation, this step is replaced by an Adam optimizer using an SPSA gradient estimator. Because SPSA estimates the gradient using only two perturbed objective evaluations per iteration, it is better suited to shot-based variational optimization and leads to a substantially lighter update step. Despite this reduction in computational cost, the numerical results showed that the SPSA-based update recovered the same solution quality as the previous finite-difference Adam optimizer.

The implementation also removes repeated work around the optimizer itself. For a fixed circuit depth, the objective evaluator is constructed once and reused across all optimization runs at that depth, rather than being rebuilt repeatedly for each run. Similarly, when a classical reference mode is requested and the instance is small enough, the classical baseline is computed once during the comparison call and reused across all modes, rather than recomputed repeatedly. In addition, lighter default settings are used for development and testing, including fewer training shots, a single averaging pass in the objective evaluator, a modest number of optimization iterations, and fewer random starts and warm-start perturbations. These choices allow experiments to be executed without unnecessarily heavy solver configurations.

\subsection{Improving Runtime Efficiency (Throughput)}

A final group of improvements targets repeated circuit execution under a fixed compiled circuit. During variational parameter optimization, the framework often needs to compute the objective value for several different parameter vectors while the QUBO problem, circuit depth, variable allocation, and execution mode remain unchanged. In this situation, the underlying quantum circuit structure is the same, and only the numerical values of the variational parameters differ. Instead of evaluating each parameter vector through a separate circuit, the implementation reuses the same transpiled circuit and inserts the variational parameter values for multiple evaluations under the same circuit structure. These evaluations are then performed together as a batch in a single execution request. This reduces the overhead of repeated execution and makes objective evaluation more efficient when several parameter vectors must be tested under the same compiled circuit. It also improves the efficiency of repeated circuit execution during variational optimization, since the same compiled circuit can be reused across multiple evaluations without additional reconstruction or retranspilation. In practice, batching is especially useful for steps that require multiple objective evaluations within the same circuit structure, such as those used in SPSA-based updates.

In addition, the framework parallelizes independent optimization runs that originate from different initial parameter vectors. Since these runs do not depend on one another, they can be executed concurrently and compared only after completion. This reduces wall-clock time for the overall multi-start search without changing the optimization logic or the final selection criteria.
Taken together, batching and parallel execution improve practical throughput and make the solver more suitable for repeated comparative studies across monolithic and distributed QAOA modes.

It is worth noting that these optimizations do not change the mathematical problem the framework solves. Instead, they reduce redundant computation around the same QUBO objective, the same variational parameter search, and the same final evaluation procedure. Their role is therefore implementation-oriented rather than algorithmic, enabling the framework to run faster and be more scalable while preserving consistency across all supported execution modes.

\section{Simulation Results and Discussions}

We evaluate the proposed framework from three perspectives: software usability, solution accuracy, and simulator runtime. First, the graphical interface is demonstrated to show how users can configure QUBO instances, select solver modes, and inspect output metrics through an interactive dashboard. Second, the numerical tests verify that the quantum-based modes recover solutions consistent with the classical reference on the tested QUBO instances. Third, the runtime analysis examines how the runtime changes across successive implementation stages, from the initial baseline implementation to the fully optimized code. The simulations are performed on QUBO instances with 6, 12, 15, and 20 binary variables. For each size, multiple distributed configurations are tested by varying the number of QPUs, including 2, 3, 4, and 5 QPUs, the capacity assigned to each QPU, and the variable-allocation strategy. The algorithmic settings are also varied, including the number of training shots for circuit-parameter optimization, the number of optimizer iterations, and the number of final sampling shots. These combinations result in approximately 100 simulation cases, allowing the framework to be evaluated in terms of recovered bitstring, objective value, optimality gap, solution consistency across modes, and runtime performance.

\subsection{Graphical Solver Interface and Reproducibility Demonstration}

Before presenting the numerical benchmark results, we first demonstrate the graphical interface developed for the proposed DQAOA-QUBO simulator. The purpose of this interface is to make the solver workflow accessible and reproducible without requiring users to modify Python scripts directly. The interface is implemented using Streamlit and wraps the same backend solver functions used in the Python package. Therefore, the graphical layer does not change the underlying optimization method; instead, it provides an interactive front end for QUBO input, solver configuration, execution control, and result visualization.

\begin{figure}[H]
\centering

\begin{subfigure}{0.9\linewidth}
\centering
\includegraphics[width=\linewidth]{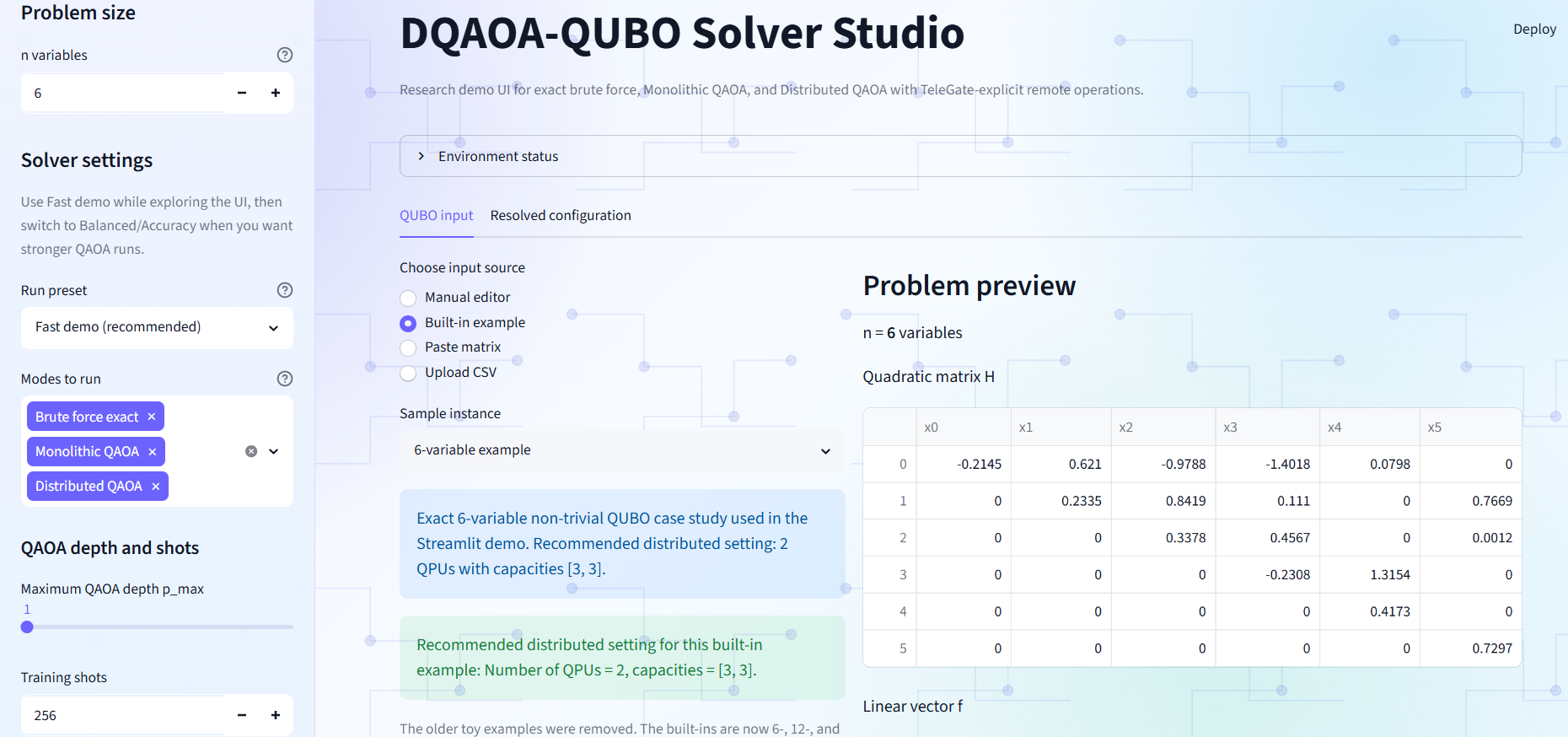}
\caption{Problem setup and built-in 6-variable QUBO preview.}
\label{fig:ui-demo-input}
\end{subfigure}

\vspace{0.5em}

\begin{subfigure}{0.9\linewidth}
\centering
\includegraphics[width=\linewidth]{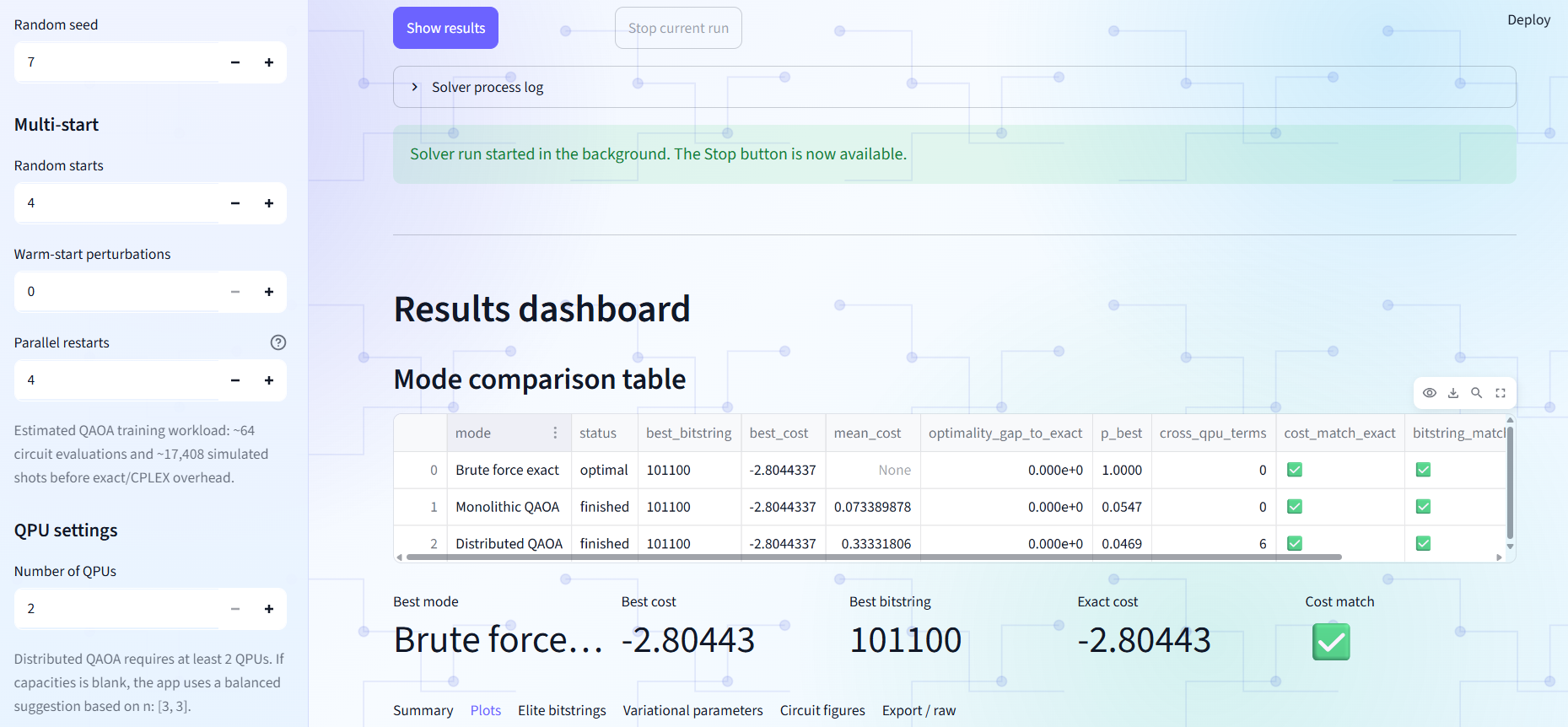}
\caption{Results dashboard showing mode comparison and solution accuracy.}
\label{fig:ui-demo-results}
\end{subfigure}

\vspace{0.5em}

\begin{subfigure}{0.9\linewidth}
\centering
\includegraphics[width=\linewidth]{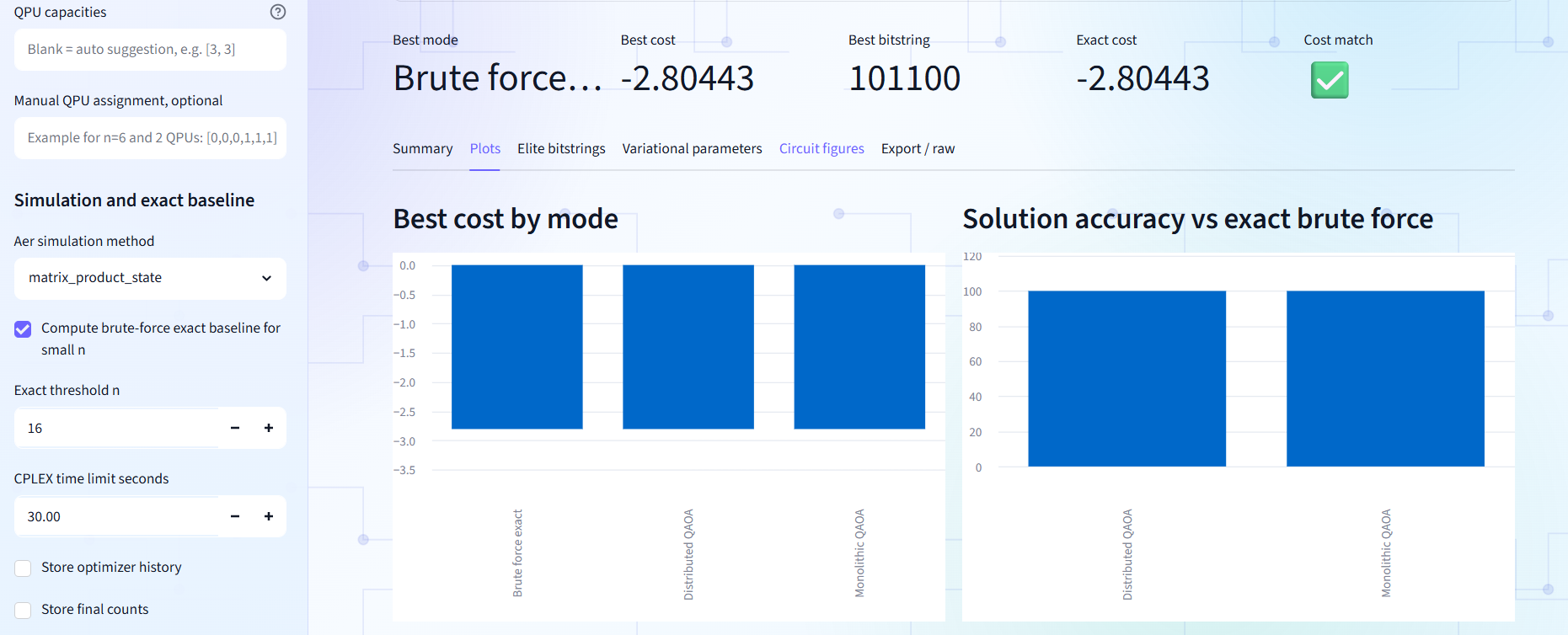}
\caption{Results dashboard showing output plots.}
\label{fig}
\end{subfigure}

\caption{Streamlit-based graphical user interface of the DQAOA-QUBO Solver Studio. The interface allows users to configure QUBO inputs, select solver modes, adjust QAOA and QPU settings, and visualize the resulting solution-quality metrics.}
\label{fig:streamlit-ui-demo}
\end{figure}

Fig.~\ref{fig:streamlit-ui-demo} shows the DQAOA-QUBO Solver Studio applied to the built-in 6-variable QUBO example. The left sidebar contains the main user-configurable settings, including the number of variables, selected solver modes, QAOA depth, training and final sampling shots, optimizer settings, multi-start settings, number of QPUs, and QPU capacities. The main panel displays the QUBO problem preview, including the quadratic matrix, linear vector, constant offset, and basic problem statistics. This allows the user to verify the problem instance before launching the solver.

After execution, the results dashboard reports the comparison among brute force, monolithic QAOA, and distributed QAOA. For the 6-variable built-in example, all selected modes recover the same best bitstring, $101100$, and the same best cost, approximately $-2.80443$. The dashboard also reports solution accuracy with respect to the brute-force reference, top-$k$ elite sampled bitstrings, cross-QPU terms, and final optimized variational parameters for the QAOA-based modes. This interface therefore provides a practical way to reproduce the solver workflow, inspect solution-quality metrics, and compare monolithic and distributed execution modes under a common input and configuration setting.

\subsection{Engineering Application: Power Generation Unit Commitment}

To demonstrate the use of the developed DQAOA package in an engineering application, the solver was also integrated into a unit commitment (UC) problem in power systems. UC is a mixed-integer optimization problem in which binary variables determine the on/off status of generating units and continuous variables determine their dispatch levels. In this test, the UC problem was solved using a three-block ADMM framework, where the continuous dispatch block was handled classically and the binary commitment block was converted into a QUBO and solved using the developed DQAOA package. This setup allows the same UC binary block to be solved using brute force, monolithic QAOA, and distributed QAOA modes under a consistent interface. Related background on distributed variational quantum methods for UC can be found in~\cite{hasanzadeh2026distributed}.

The UC test case includes five generating units over three time periods, resulting in 15 binary commitment variables. Table~\ref{tab:uc_application_results} reports the optimal bitstring and cost obtained by each solver mode. All three modes recover the same best commitment bitstring, $001000100101111$, with the same total operating cost of $12678.45$, which confirms that the monolithic and distributed QAOA modes preserve solution consistency with the brute force reference when the DQAOA package is used as the QUBO solver inside the ADMM-based UC workflow.

\begin{table}[H]
\centering
\caption{UC application results using the DQAOA package as the QUBO solver}
\label{tab:uc_application_results}
\begin{tabular}{lcc}
\hline
\textbf{Mode} & \textbf{Best bitstring} & \textbf{Best cost} \\
\hline
Brute force       & 001000100101111 & 12678.45 \\
Monolithic QAOA   & 001000100101111 & 12678.45 \\
Distributed QAOA  & 001000100101111 & 12678.45 \\
\hline
\end{tabular}
\end{table}

\begin{figure}[H]
\centering

\begin{subfigure}{0.49\linewidth}
\centering
\includegraphics[width=\linewidth]{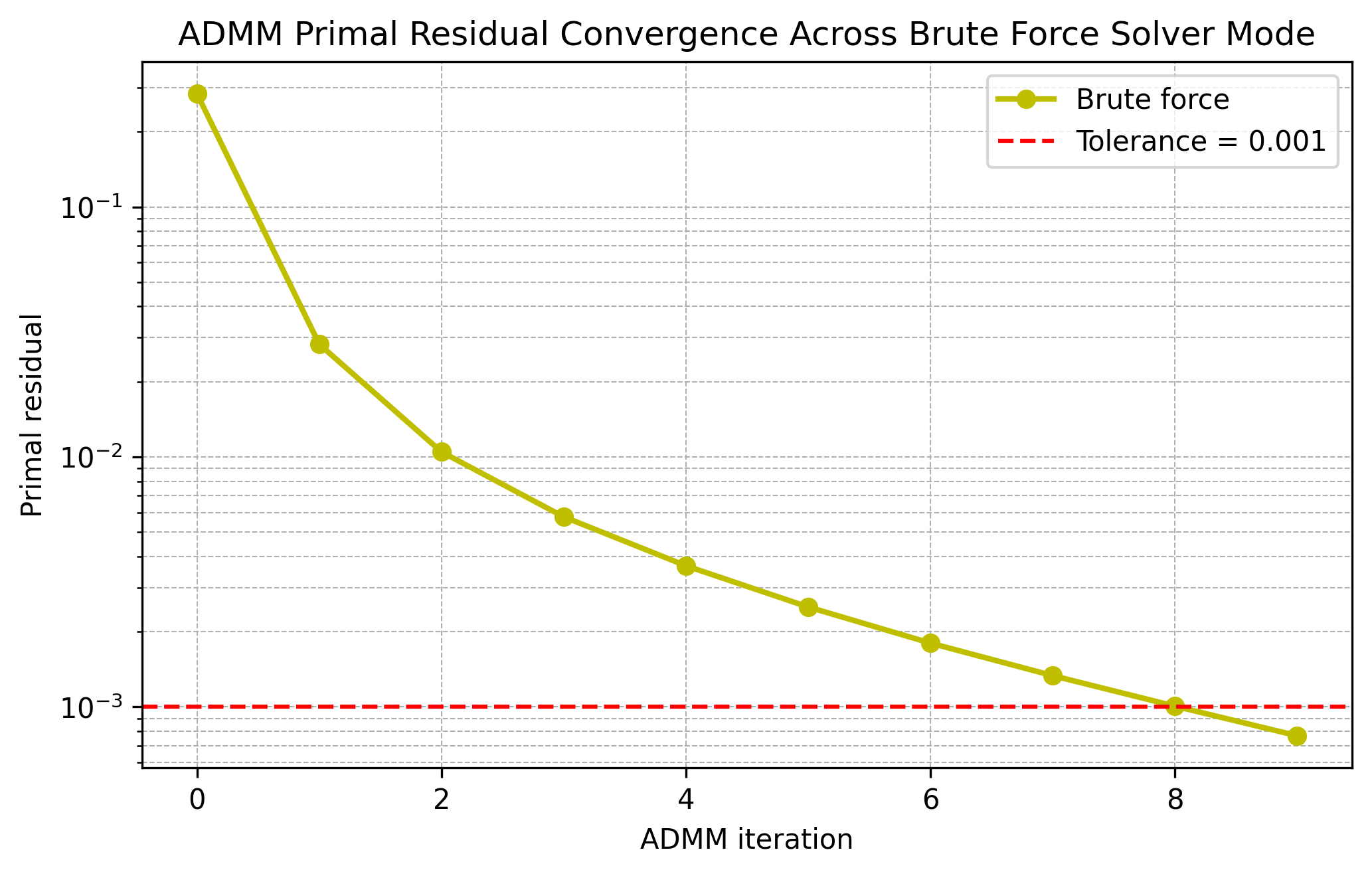}
\caption{Brute-force}
\label{fig:uc_residual_bf}
\end{subfigure}
\hfill
\begin{subfigure}{0.49\linewidth}
\centering
\includegraphics[width=\linewidth]{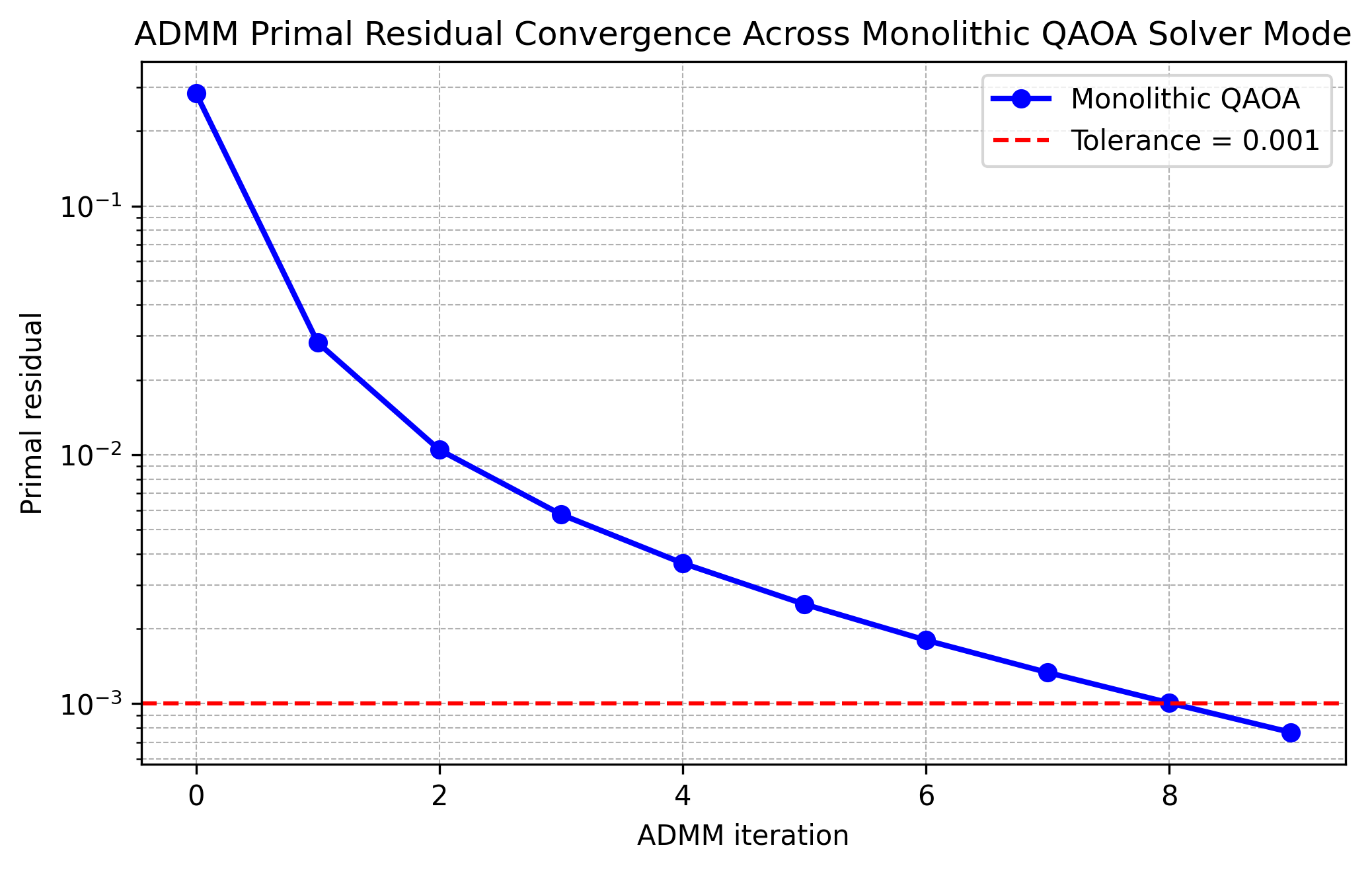}
\caption{Monolithic QAOA}
\label{fig:uc_residual_mono}
\end{subfigure}

\vspace{0.6em}

\begin{subfigure}{0.49\linewidth}
\centering
\includegraphics[width=\linewidth]{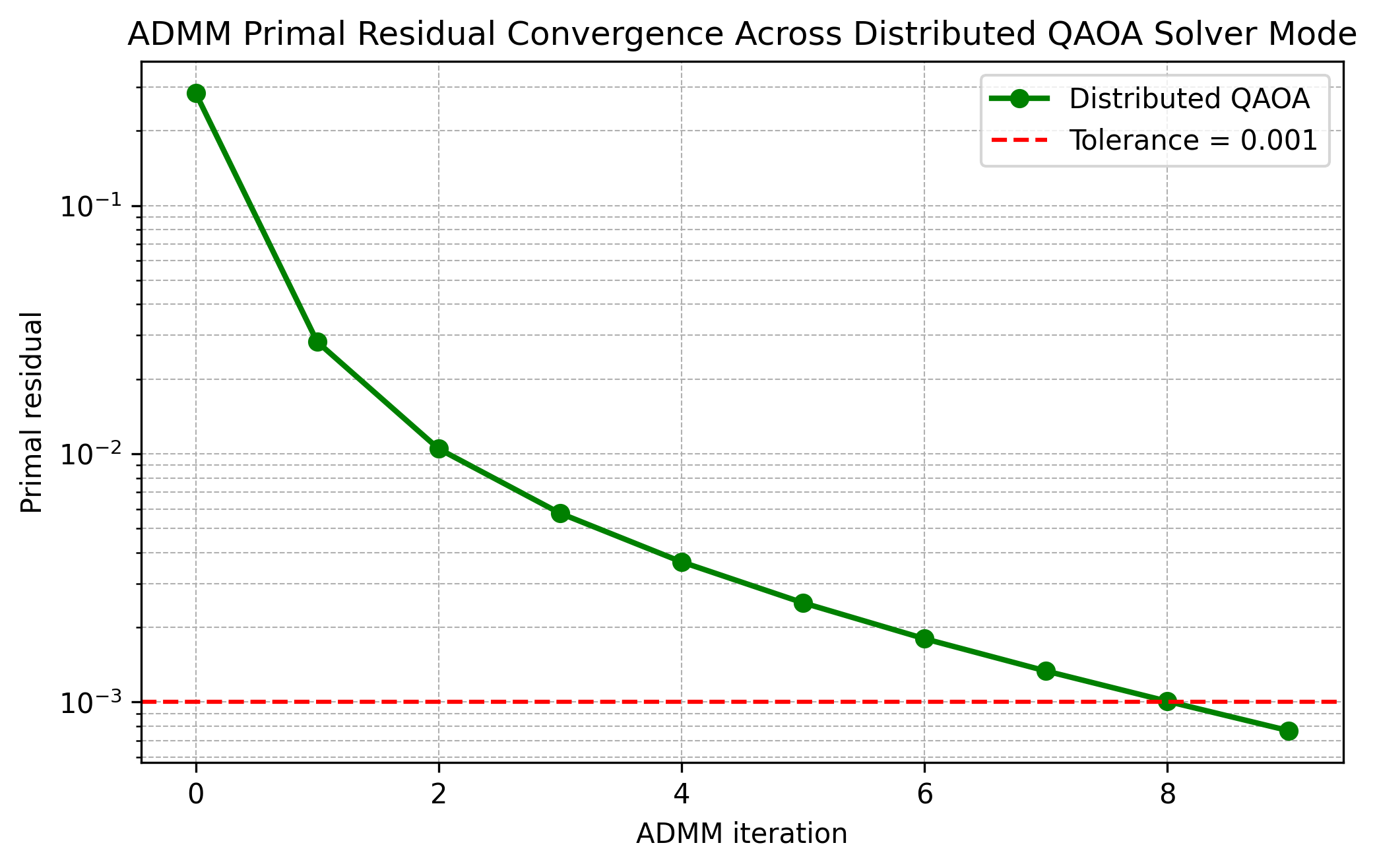}
\caption{Distributed QAOA}
\label{fig:uc_residual_dist}
\end{subfigure}
\hfill
\begin{subfigure}{0.49\linewidth}
\centering
\includegraphics[width=\linewidth]{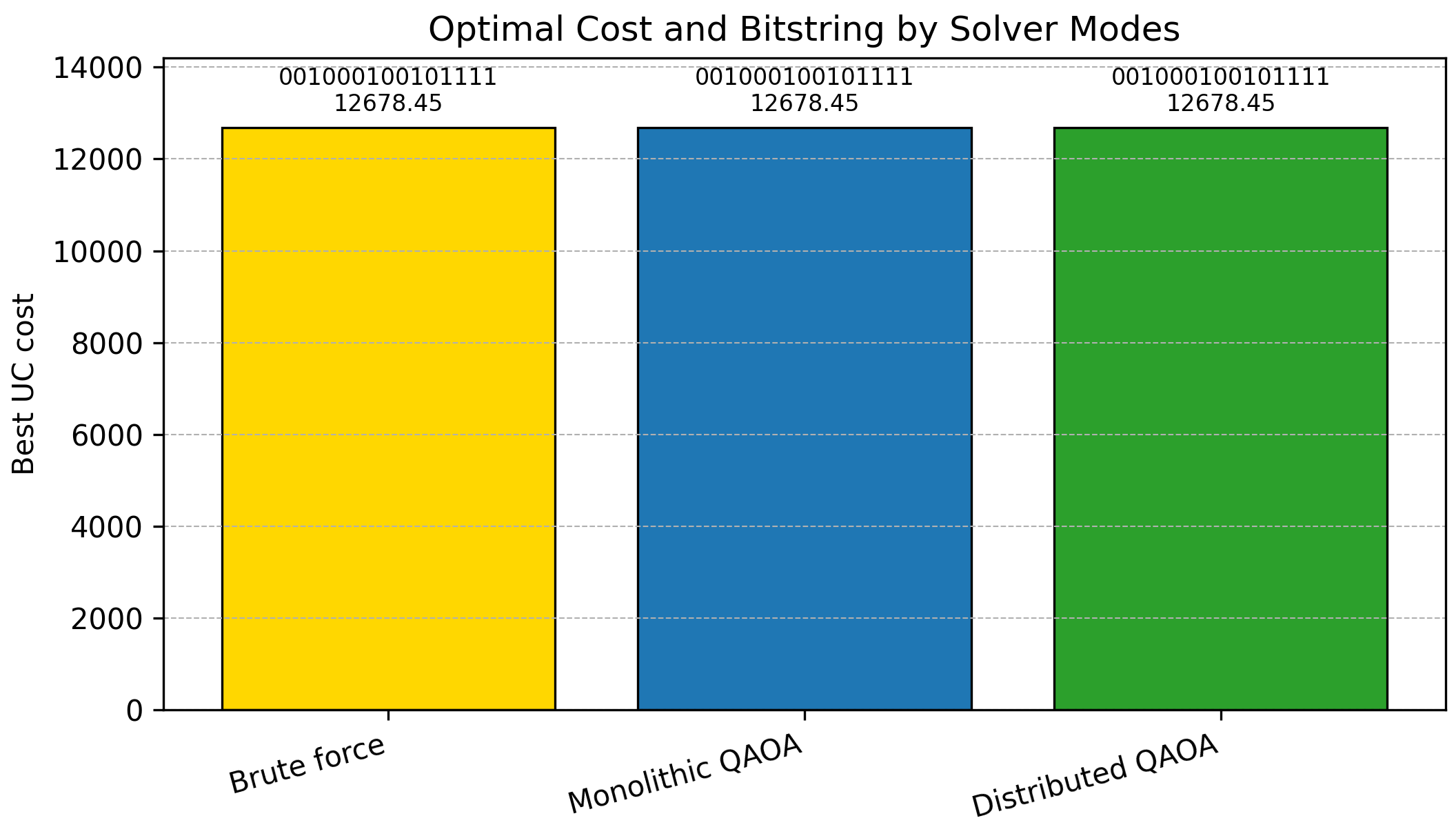}
\caption{Optimal cost and bitstring}
\label{fig:uc_best_cost_bitstring}
\end{subfigure}

\caption{UC application results using the DQAOA package as the QUBO solver inside the three-block ADMM framework. Panels (a)--(c) show the ADMM primal residual convergence for brute force, monolithic QAOA, and distributed QAOA, respectively. Panel (d) compares the optimal bitstring and operating cost across the three solver modes.}
\label{fig:uc_application_summary}
\end{figure}

The ADMM primal residual convergence and the recovered best solution are summarized in Fig.~\ref{fig:uc_application_summary}. In all three modes, the residual decreases below the adjusted tolerance of $10^{-3}$. In addition, all modes recover the same best commitment bitstring, $001000100101111$, with the same operating cost of $12678.45$.

\subsection{Runtime Comparison in Implementation Stages}

The runtime comparison is organized across four software stages. The first stage is the baseline implementation, in which no dedicated runtime optimization is applied. The second stage introduces parameterized circuit reuse: a circuit for a fixed problem, depth, allocation, and execution mode is constructed and transpiled once, then reused by binding updated parameter values. The third stage further reduces runtime by replacing the original finite difference gradient calculation, which perturbs one parameter at a time, with an SPSA-based parameter update optimizer. It also avoids rebuilding the objective evaluator repeatedly at a fixed circuit depth and avoids recomputing the classical reference separately for each mode. The final stage adds batched circuit evaluations and parallel restart execution. These stages correspond directly to the implementation changes described in the previous section.

\subsection{Solution Accuracy Across Modes}

Before comparing runtime, it is important to verify that the classical reference, monolithic QAOA mode, and distributed QAOA mode produce consistent solutions on the tested QUBO instances. Tables~\ref{tab:accuracy_ex1} and \ref{tab:accuracy_ex2} report the best bitstring and best cost for the 6-variable and 12-variable case studies, respectively. In both cases, the quantum modes recover the same optimal bitstring and cost as the classical reference. This confirms that the runtime comparisons reported later are carried out between modes that maintain consistent solution quality.

\begin{table}[t]
\centering
\caption{Best bitstring and best cost for the 6-variable case study.}
\label{tab:accuracy_ex1}
\begin{tabular}{lcc}
\hline
\textbf{Mode} & \textbf{Best bitstring} & \textbf{Best cost} \\
\hline
Brute force       & 101100 & -2.8 \\
Monolithic QAOA   & 101100 & -2.8 \\
Distributed QAOA  & 101100 & -2.8 \\
\hline
\end{tabular}
\end{table}

\begin{table}[t]
\centering
\caption{Best bitstring and best cost for the 12-variable case study.}
\label{tab:accuracy_ex2}
\begin{tabular}{lcc}
\hline
\textbf{Mode} & \textbf{Best bitstring} & \textbf{Best cost} \\
\hline
Brute force       & 011110010110 & -2.6 \\
Monolithic QAOA   & 011110010110 & -2.6 \\
Distributed QAOA  & 011110010110 & -2.6 \\
\hline
\end{tabular}
\end{table}

\subsection{6-Variable Case Study}

\begin{figure}[H]
  \centering

  \begin{subfigure}{0.75\linewidth}
    \centering
    \includegraphics[width=\linewidth]{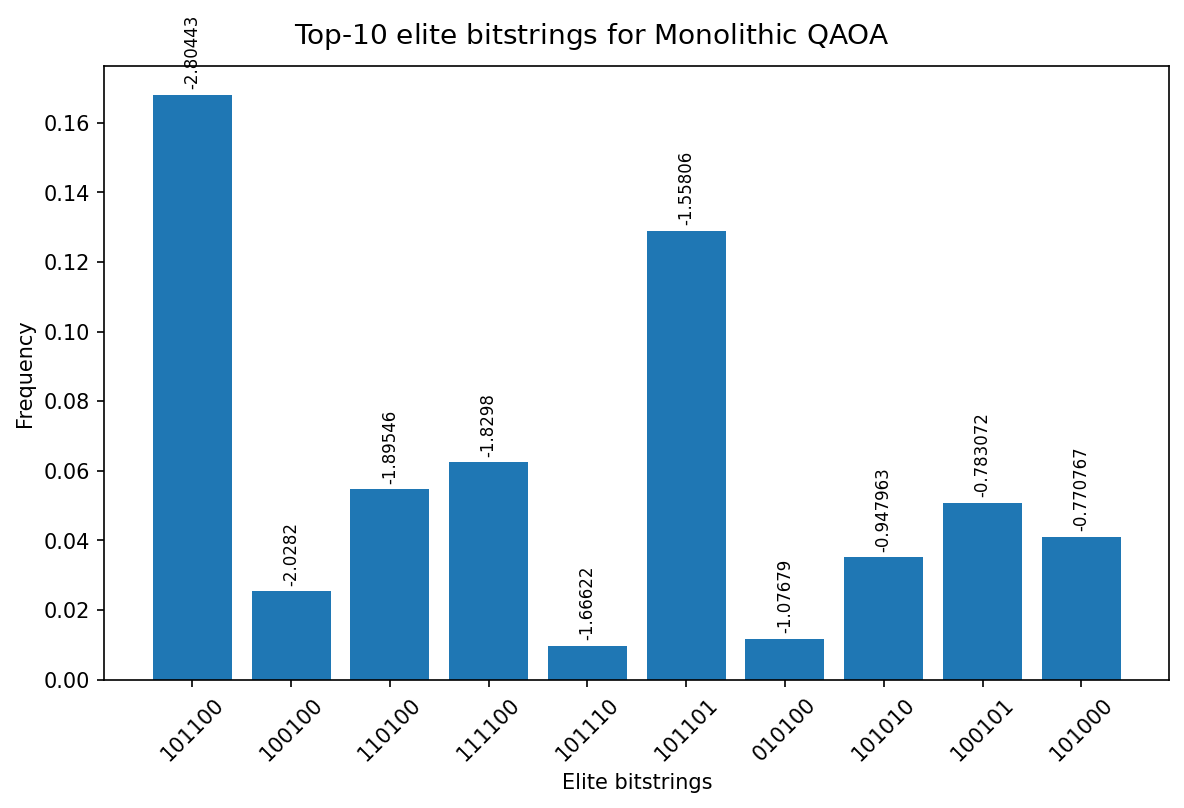}
    \caption{Monolithic QAOA.}
    \label{fig:elite-6var-mono}
  \end{subfigure}

  \vspace{0.5em}

  \begin{subfigure}{0.75\linewidth}
    \centering
    \includegraphics[width=\linewidth]{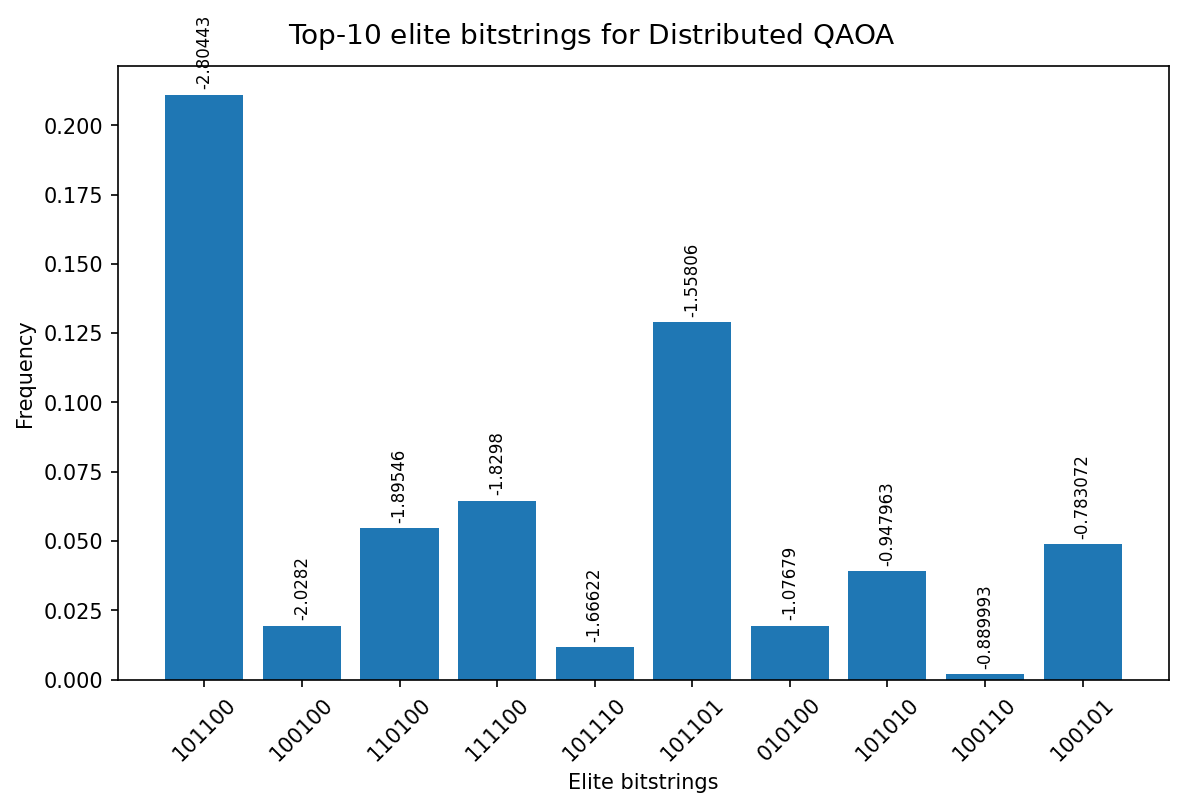}
    \caption{Distributed QAOA.}
    \label{fig:elite-6var-dist}
  \end{subfigure}

  \caption{Top-10 elite sampled bitstrings for the 6-variable QUBO case study.}
  \label{fig:elite-6var}
\end{figure}

Before analyzing the runtime behavior, the solution consistency of the monolithic and distributed QAOA modes is first examined through the final sampled distributions. Figure~\ref{fig:elite-6var} reports the top-10 elite sampled bitstrings obtained from the final shot-based evaluation of both modes. The bars represent the sampled frequencies of the elite bitstrings, while the values above the bars show the corresponding QUBO costs. As shown in Fig.~\ref{fig:elite-6var}, both execution modes identify the same lowest-cost sampled bitstring, \(101100\), with the same QUBO cost of \(-2.80443\). This confirms that the distributed QAOA implementation preserves the best sampled solution obtained by the monolithic QAOA mode for this small instance. The remaining elite bitstrings also include similar low-cost candidates. However, their frequencies are not identical because the two modes use different circuit realizations and the final distributions are estimated from a finite number of shots. These results provide the solution-consistency basis for the runtime comparison presented next.

After confirming solution consistency across modes, Table~\ref{tab:runtime_ex1} reports the runtime progression for the 6-variable QUBO instance. The brute-force reference solves this small instance almost instantaneously, whereas the quantum modes require a much longer runtime even in the monolithic setting. This is expected because the two approaches (brute-force and quantum-based modes) perform fundamentally different amounts of work. In the brute-force case, only \(2^6=64\) candidate bitstrings must be evaluated, and the cost of each bitstring can be computed directly. In contrast, the quantum solver must perform a variational optimization procedure in which the circuit is repeatedly executed for different parameter values, across multiple circuit depths and starting points, and each objective value is estimated from measurement samples. As a result, even for a small benchmark instance, the total computational effort of monolithic QAOA is much larger than that of the classical reference.

The first implementation refinement already yields a clear runtime reduction. In the monolithic QAOA mode, the runtime decreases from 46.54~seconds in the baseline code to 6.87~seconds after parameterized circuit reuse is introduced. In distributed QAOA, the runtime decreases from 224 to 147 seconds. The next stage further reduces runtime to 5 seconds in the monolithic QAOA mode and 68 seconds in the distributed QAOA mode. The fully optimized implementation gives the best distributed runtime of 62 seconds, while keeping the monolithic runtime in the same low range (4 seconds).

Distributed QAOA remains slower than monolithic QAOA at every stage. This is not because the two modes solve different optimization problems. In both cases, the same QUBO instance is used and the same expected QUBO cost is minimized. The difference lies in the circuit implementation. In the monolithic mode, all required interactions are executed locally within one QPU. In DQAOA mode, interactions that cross QPU boundaries must be implemented via explicit remote operations, which increase circuit complexity and runtime. In the present framework, this realization uses auxiliary communication qubits, mid-circuit measurements, resets, and classical feedforward corrections. As a result, the distributed circuit is structurally more complex than the monolithic circuit, and its runtime increases accordingly. This behavior is consistent with the role of local and cross-QPU couplings introduced in the DQAOA framework and with the communication-related overhead discussed in the runtime optimization section. 

\begin{table}[t]
\centering
\caption{Runtime comparison for the 6-variable case study (seconds).}
\label{tab:runtime_ex1}
\begin{tabular}{lccc}
\hline
\textbf{Implementation stage} & \textbf{Brute-force} & \textbf{Monolithic QAOA} & \textbf{Distributed QAOA} \\
\hline
Baseline              & 0.00032 & 46.54 & 223.54 \\
First-level optimized & 0.00030 & 6.87  & 147.26 \\
Intermediate optimized& 0.00030 & 4.55  & 68.44 \\
Fully optimized       & 0.00022 & 4.44  & 62.49 \\
\hline
\end{tabular}
\end{table}

\subsection{12-Variable Case Study}

\begin{figure}[H]
\centering

\begin{subfigure}{0.75\linewidth}
\centering
\includegraphics[width=\linewidth]{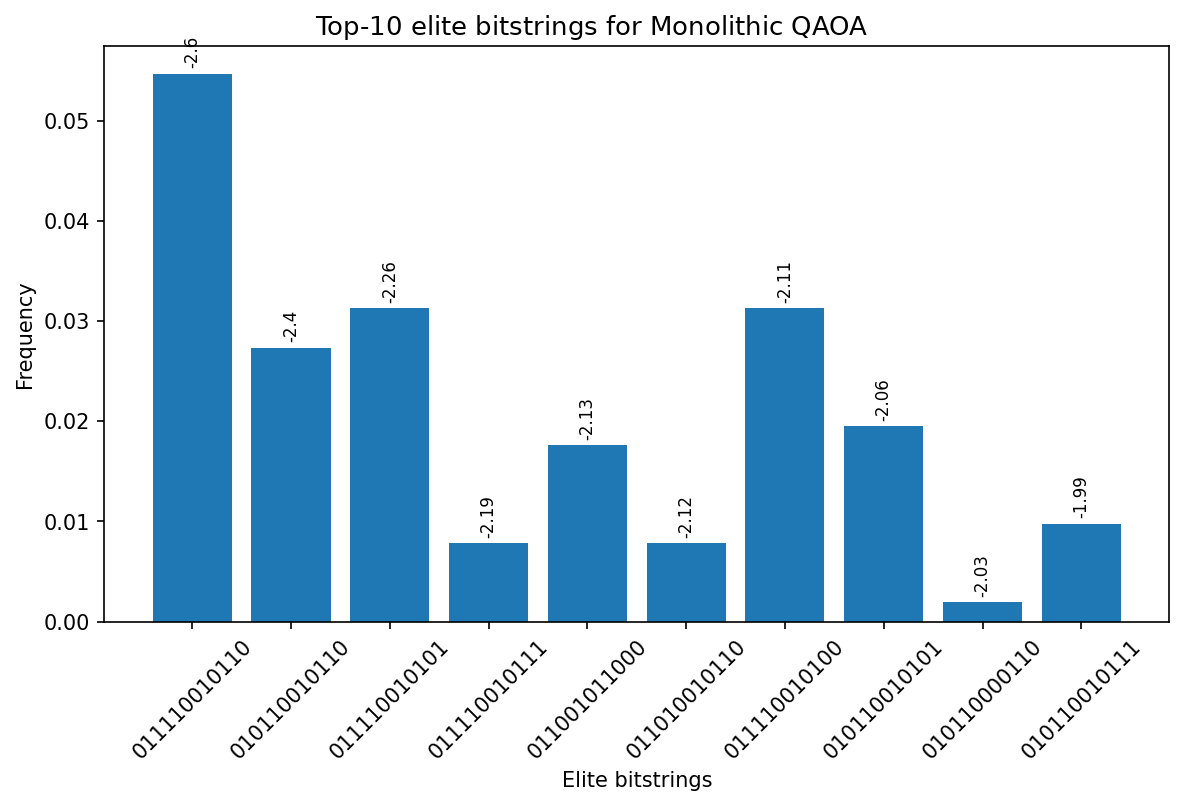}
\caption{Monolithic QAOA.}
\label{fig:elite-12var-mono}
\end{subfigure}

\vspace{0.5em}

\begin{subfigure}{0.75\linewidth}
\centering
\includegraphics[width=\linewidth]{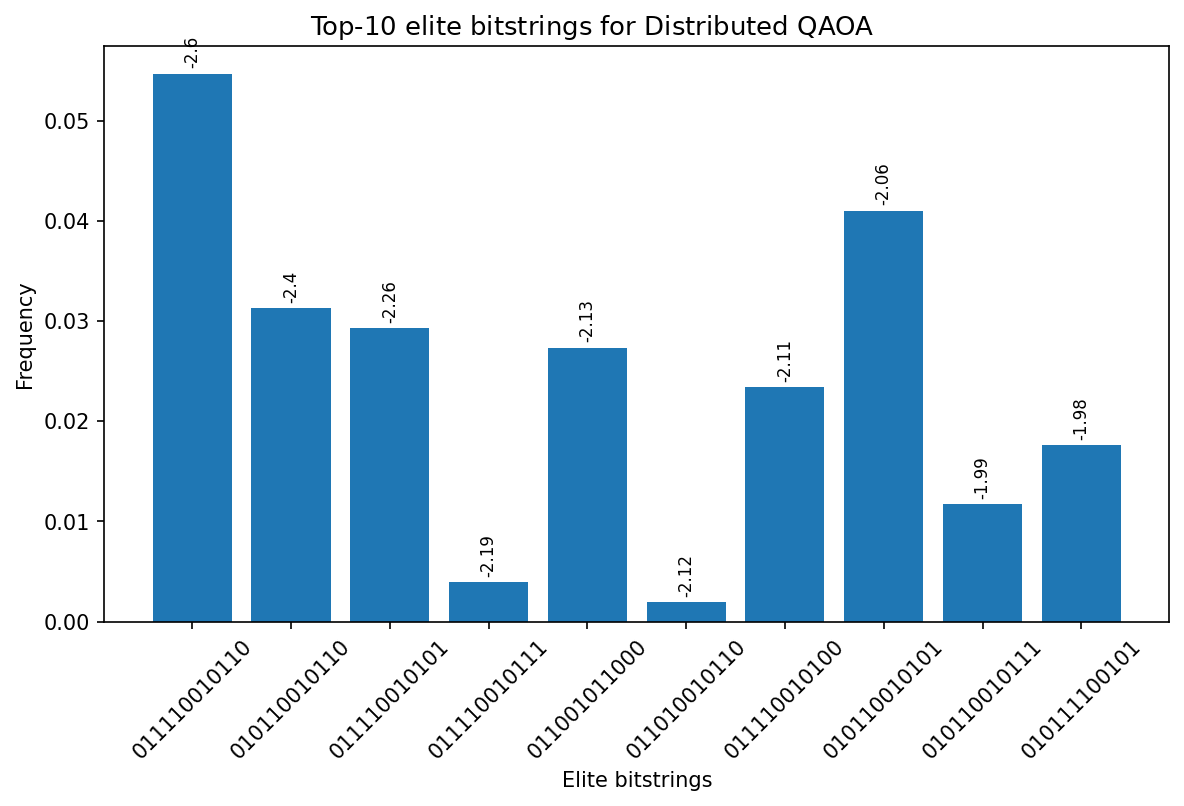}
\caption{Distributed QAOA.}
\label{fig:elite-12var-dist}
\end{subfigure}

\caption{Top-10 elite sampled bitstrings for the 12-variable QUBO case study.}
\label{fig:elite-12var}
\end{figure}

The same accuracy and consistency are observed in the 12-variable case, as reported in Fig.~\ref{fig:elite-12var}. However, as reported in Table~\ref{tab:runtime_ex2}, the differences in runtime become more noticeable. As in the 6-variable example, the brute-force reference remains much faster than the quantum modes, while the DQAOA mode remains slower than the monolithic QAOA mode. In this case, however, the larger problem size makes the effect of repeated circuit evaluations and cross-QPU operations more evident in the total runtime.

From an implementation standpoint, the benefits of the optimization strategies are clearer in this example. In the baseline implementation, the runtime is 456.63~seconds in the monolithic QAOA mode and 18929.12~seconds in the Distributed QAOA mode. After introducing parameterized circuit reuse, these values decrease to 110.47~seconds and 9684.59~seconds, respectively. The intermediate implementation reduces them further to 13.35~seconds and 1450.74~seconds. The fully optimized version reaches 11.52~seconds in the monolithic QAOA mode and 1224.64~seconds in the Distributed QAOA mode. 

These results show that runtime improvements become increasingly important as the problem size grows. They also show that, even after software overhead is reduced, the DQAOA mode takes longer than the monolithic QAOA mode because the cost of multi-QPU execution becomes increasingly dominant. Thus, for the 12-variable case, the runtime comparison confirms the effectiveness of the proposed implementation refinements and shows that multi-QPU execution in the distributed QAOA mode still requires substantially more runtime than the monolithic QAOA mode.

\begin{table}[t]
\centering
\caption{Runtime comparison for the 12-variable case study (seconds).}
\label{tab:runtime_ex2}
\begin{tabular}{lccc}
\hline
\textbf{Implementation stage} & \textbf{Brute force} & \textbf{Monolithic QAOA} & \textbf{Distributed QAOA} \\
\hline
Baseline              & 0.10359 & 456.63 & 18929.12 \\
First-level optimized & 0.02401 & 110.47 & 9684.59 \\
Intermediate optimized& 0.02294 & 13.35  & 1450.74 \\
Fully optimized       & 0.02067 & 11.52  & 1224.64 \\
\hline
\end{tabular}
\end{table}

\subsection{15-variable Case Study}

To further illustrate the behavior of the final code at a larger size, a 15-variable QUBO instance was also solved using the fully optimized implementation. In this case, the brute-force runtime was 0.2 seconds, the monolithic QAOA runtime was 17 seconds, and the Distributed QAOA runtime was 6083 seconds. This result follows the same pattern observed in the smaller cases. The classical reference remains much faster on a benchmark of this size, the monolithic QAOA mode incurs the cost of repeated variational circuit evaluations, and the distributed QAOA mode requires additional circuit operations to realize interactions between variables assigned to different QPUs.

Overall, the results show that the proposed runtime refinements produce a substantial practical improvement. The first stage removes the overhead of repeated circuit construction and transpilation; the second stage reduces the cost of parameter optimization and repeated setup work; and the final stage improves execution efficiency during repeated evaluations and multi-start search. At the same time, the results show that the DQAOA mode remains slower than monolithic QAOA because interactions that cross QPU boundaries require extra circuit operations that do not arise in the single-QPU implementation.

\section{Conclusion}

This work presented a Qiskit-compatible DQAOA simulator package for solving QUBO problems under both monolithic and distributed quantum execution settings. The motivation for developing this package was the lack of a general software workflow that could accept QUBO instances and consistently evaluate them across different QAOA execution modes. The proposed package supports classical reference modes, monolithic QAOA on a single QPU, and Distributed QAOA across a user-specified number of QPUs with configurable capacities. The number of logical qubits is determined by the size of the input QUBO instance, allowing users to study different problem sizes while adjusting multi-QPU configuration based on the target simulation setting.

Starting from a canonical QUBO representation, the framework maps the objective to the corresponding cost Hamiltonian, allocates variables across QPUs, identifies local and cross-QPU couplings, constructs the corresponding circuits, and evaluates the final sampled bitstrings within one consistent workflow. The work also showed that runtime optimization strategies are important for practical usability. Parameterized circuit reuse, SPSA-based parameter updates, objective reuse at fixed depth, batched evaluations, and parallel multi-start execution reduced the need for repeated software-side overhead. It helped facilitate the solution and comparison of QUBO problems within the proposed package.

Numerical results on two case studies showed that monolithic QAOA and Distributed QAOA remained consistent with classical references in terms of the best bitstring and best cost, while the staged runtime comparisons showed substantial reductions across successive implementation stages. The larger cases further showed that DQAOA remains more computationally demanding than monolithic QAOA, as cross-QPU interactions must be implemented explicitly via the TeleGate approach. The UC application further proved that the proposed package can be used as a QUBO solver module inside a larger engineering optimization framework, where brute force, monolithic QAOA, and distributed QAOA modes recover the same commitment bitstring and operating cost under the three-block ADMM workflow. The graphical user interface further improves the usability of the package by allowing users to configure QUBO instances, execute selected solver modes, and visualize solution accuracy, elite sampled bitstrings, and optimized variational parameters through an interactive dashboard. The software implementation is publicly available through the \href{https://sites.google.com/site/aminkargarian/home}{RAISE LAB website} and \href{https://github.com/LSU-RAISE-LAB/}{RAISE LAB GitHub repository}, together with README documentation for installation and usage. Overall, the proposed package provides a reusable Qiskit-compatible software foundation for studying QUBO-based quantum optimization across single-QPU and multi-QPU execution settings.

\bibliographystyle{unsrt}
\bibliography{references}
\end{document}